\documentclass[reprint,
superscriptaddress,
amsmath,amssymb, longbibliography,
aps, prb,
floatfix]{revtex4-2}

\usepackage{bm}
\usepackage[acronym,toc,automake=immediate,nonumberlist,nopostdot,nogroupskip]{glossaries}
\usepackage[hidelinks]{hyperref}
\hypersetup{breaklinks=true}
\hypersetup{
    colorlinks=true,
    linkcolor=blue,
    filecolor=magenta,      
    urlcolor=blue,
    pdftitle={Electric-Field Driven Nuclear Dynamics of Liquids and Solids from a Multi-Valued Machine-Learned Dipolar Model},
    pdfpagemode=FullScreen,
}
\usepackage{xspace}
\usepackage{xcolor}
\usepackage{mathtools,amssymb,amscd,amsthm,amsmath,graphicx,color,bbm,bm,amsbsy,amsfonts}
\usepackage{braket}
\usepackage{booktabs}
\usepackage[inline]{enumitem}
\usepackage[normalem]{ulem}

\DeclareMathAlphabet{\mathbbold}{U}{bbold}{m}{n}

\usepackage{array}
\newcolumntype{R}[1]{>{\raggedleft\arraybackslash}p{#1}}
\newcolumntype{L}[1]{>{\raggedright\arraybackslash}p{#1}}
\newcolumntype{C}[1]{>{\centering\arraybackslash}p{#1}}

\newcommand{\rev}[1]{{\textcolor{black}{#1}}}

\makeglossaries
\newacronym{BO}{BO}{Born-Oppenheimer}
\newacronym{DFT}{DFT}{density-functional theory}
\newacronym{PES}{PES}{potential energy surface}
\newacronym{BEC}{BEC}{Born effective charges}
\newacronym{EDA}{EDA}{electric dipole approximation}
\newacronym{ML}{ML}{machine-learning}
\newacronym{ENN}{ENN}{equivariant neural network}
\newacronym{FPS}{FPS}{farthest point sampling}
\newacronym{mlip}{MLIP}{machine-learning interatomic potential}
\newacronym{PIMD}{PIMD}{path integral molecular dynamics}
\newacronym{MSD}{MSD}{mean square displacement}

\newlist{mylist}{enumerate*}{1}
\setlist[mylist]{label=(\roman*)}
\setlist[itemize]{itemsep=0pt}



\setcitestyle{doi=false}
\setcitestyle{url=false}
\setcitestyle{eprint=false}

\newcommand{\editor}[2]{%
  \expandafter\newcommand\csname #1note\endcsname[1]{%
    \textcolor{#2}{(\textbf{#1:} ##1)}}%
  \expandafter\newcommand\csname #1cancel\endcsname[1]{%
    \textcolor{#2}{\sout{##1}}}%
  \expandafter\newcommand\csname #1change\endcsname[2]{%
    \textcolor{#2}{\sout{##1} ##2}}%
  \newenvironment{#1text}{\color{#2}}{\color{black}}
  \expandafter\newcommand\csname #1add\endcsname[1]{%
    \textcolor{#2}{##1}}
}%
\editor{ES}{blue}

\begin{document}

    \title{Electric-Field Driven Nuclear Dynamics of Liquids and Solids from a Multi-Valued Machine-Learned Dipolar Model}
    
    \author{Elia Stocco}
    \affiliation{MPI for the Structure and Dynamics of Matter, Luruper Chaussee 149, 22761 Hamburg, Germany}
    
    \author{Christian Carbogno}
    \affiliation{Theory Department, Fritz Haber Institute of the MPS, Faradayweg 4-6, 14195 Berlin, Germany}

    \author{Mariana Rossi}
    \email{mariana.rossi@mpsd.mpg.de}
     \affiliation{MPI for the Structure and Dynamics of Matter, Luruper Chaussee 149, 22761 Hamburg, Germany}
    
    \begin{abstract}
        \noindent
The driving of vibrational motion by external electric fields is a topic of continued interest, due to the possibility of assessing new or metastable material phases with desirable properties.
Here, we combine {\it ab initio} molecular dynamics within the electric-dipole approximation with machine-learning neural networks (NNs) to develop a general, efficient and accurate method to perform electric-field-driven nuclear dynamics for molecules, solids, and liquids.
We train equivariant and autodifferentiable NNs for the interatomic potential and the dipole, modifying the \rev{model infrastructure} to account for the multi-valued nature of the latter in periodic systems.
We showcase the method by addressing property modifications induced by electric field interactions in a polar liquid and a polar solid from nanosecond-long molecular dynamics simulations with quantum-mechanical accuracy. For liquid water, we present a calculation of the dielectric function in the GHz to THz range and the electrofreezing transition, showing that nuclear quantum effects enhance this phenomenon. For the ferroelectric perovskite LiNbO$_3$, we simulate the ferroelectric to paraelectric phase transition and the non-equilibrium dynamics of driven phonon modes related to the polarization switching mechanisms, showing that a full polarization switch is not achieved in the simulations.
    \end{abstract}
   
    \maketitle

    \section{Introduction \label{sec:intro}}

        The interaction of matter with static and dynamical external electric fields plays a fundamental role in understanding the behavior of atoms, molecules, and complex materials. 
        This interaction is exploited in fields as diverse as semiconductor nanotechnology~\cite{Weintrub+natcom2022} and enzymatic reactions~\cite{FriedBoxer+arb2017}. 
        In particular, there is a rising interest in tuning static and low-frequency dielectric fields to drive reversible phase transitions of materials~\cite{SALEN20191, monacelli, knbo_2025, NicolettiCavalleri2016}, with the goal of controlling mechanical properties, increasing the efficiency of ionic conductors and capacitors, or making better energy or memory storage devices~\cite{Weintrub+natcom2022, FriedBoxer+arb2017}. 
        The large spatial extent and, depending on the phenomenon, the time-scale of many nanoseconds involved in the nuclear dynamics that define the response of a material to the application of such electric fields has been a long-standing challenge for quantum-mechanical first-principles atomistic simulations. While the physics of the matter-field coupling is known and appropriate simulation techniques already exist within the context of quantum-mechanical theories~\cite{Stengel2009, UmariPasquarello2002,PhysRevLett.89.117602}, 
        simplified and non-transferable models~\cite{SALEN20191} have often been used for describing these quantum phenomena at larger system sizes and longer time scales in complex systems, especially when targeting non-equilibrium dynamics.

        It is clear that the rapidly evolving techniques of \gls{ML} in atomistic simulations can be applied to this problem. \gls{ML} models have already been proven to succeed in the prediction of  energies, forces and electronic-structure properties of materials \cite{Kulik_2022,guideML} with an accuracy comparable to that of the underlying quantum electronic-structure method, but at a small fraction of the cost and often with a strongly reduced scaling with system size. A fundamental quantity to describe the interaction of matter with electric fields is the dipole and its derivative with respect to nuclear displacements. Several works have already proposed ML-based models for these quantities~\cite{schmiedmayer2024derivative, Gigli2022, Jana2024, Litman2023, Shimizu, falletta2024unified, Knijff_2021, VeitCeriottiJCP,  Schienbein2024, PhysRevB.102.041121, gastegger2017machine, Shimizu}. While these approaches are appropriate and successful in the investigated examples,~i.e.,~close to structural equilibria, they can become inaccurate when large displacements take place, like in ionic diffusion or in non-equilibrium dynamics. As we will show in this work, an additional term arising from the multi-valued nature of the dipole in periodic systems sensitively influences the applicability of the \gls{ML} model in these cases. 

        By incorporating the multi-valued nature of the dipole in the ML model, we achieve a general framework that:
        \begin{mylist}
            \item is generally applicable to different material classes (e.g.\ molecules, solids, liquids, and disordered systems) without relying on system-specific assumptions;
            \item keeps quantum-mechanical accuracy for electronic interactions and can include nuclear quantum effects;
            \item allows performing nanoseconds-long simulations including thousands of atoms in and out of equilibrium.
        \end{mylist}
         We showcase the capabilities of our model by addressing liquid water and the solid ferroelectric perovskite LiNbO$_3$~\cite{weis1985lithium}. We calculate nanoseconds of first-principles-quality molecular dynamics coupled to static and time-dependent electric fields for these polar materials. For water, we show simulations of the dielectric function~\cite{watericebook} spanning the GHz to the THz range and of the electrofreezing transition~\cite{electrofreezing2019,Cassone2024,Cassone2019}, finding an enhancement of this phenomenon due to nuclear quantum effects at larger field intensities. For LiNbO$_3$, we show the calculation of the ferroelectric to paraelectric phase transition~\cite{linobate1965,LNCurieTemp1,Smolenskii,linobate2008} and of the full-dimensional non-equilibrium and non-linear driving of phonon-modes in this material. For this last example, incorporating the multi-valued nature of the dipole in the ML model is essential. It shows that full-dimensional simulations can better explain the experimental observation that an ultrafast polarization switch is only transiently and partially achieved~\cite{MankowskiCavalleri2017,vonHoegen2018,depolarizationLINOBATE}. 
        
        In the following, we will explain the ML model we develop, and discuss the rich phenomenology that these simulations can address together with the novel physical insights they can bring.

    \section{Results}

        \subsection{Machine-learning Model for Dipoles}\label{sec:math}

\begin{figure}
                \centering
                \includegraphics[width=0.4\textwidth]{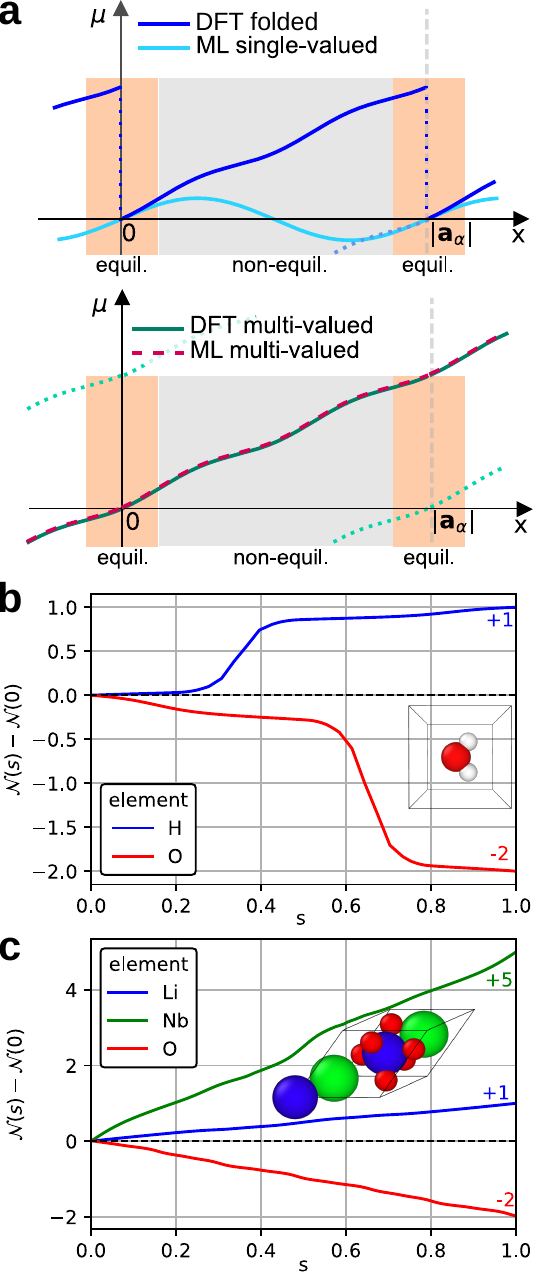}
                \caption{%
                Multi-valued dipoles and oxidation numbers.
                $\textbf{(a)}$ 
                The top panel shows a sketch of the \gls{DFT} dipole values when folded on different branches: this leads to a discontinuous function that assumes exactly the same value for the same atomic environment ($x=0$ mod $|\bm{a}_{\alpha}|$). 
                A single-valued \gls{ML} approach can be successful if the training data and the predicted configuration lie in the ``equilibrium'' region (orange).
                The bottom panel shows the \gls{DFT} dipole values lying on the same branch and leading to a continuous function. This requires a multi-valued \gls{ML} model that is able to  predict different values for the same atomic environment. This approach is also successful in the ``far from equilibrium'' region.  
                $\textbf{(b)}$ \gls{DFT} values of the projection of the dipole along a given path for a water monomer (see Eq.~\eqref{eq:oxn-compute}). The difference between the initial and final values (corresponding to equivalent configurations) correspond to the  oxidation numbers of the atoms. %
                $\textbf{(c)}$ Analogous to \textbf{b} for LiNbO$_3$. 
                }
                \label{fig:oxn}
            \end{figure}

            \rev{We start by characterizing the quantity we want to model. As further detailed in Section~\ref{sec:dipoles}, the multi-valued nature of dipoles in periodic systems has the following consequences.} If one imposes \rev{lattice periodicity} on the dipole $\bm{\mu}$ by enforcing it to assume the same value at \rev{periodically}-equivalent atomic sites, the function $\bm{\mu}(\bm{R})$ becomes discontinuous because this implies ``switching branches'' across the periodic boundaries, as shown in Fig.~\ref{fig:oxn}a. Instead, a smooth and continuous function is obtained by continuing to follow the same branch. Since the dipole then assumes different values for \rev{periodically}-equivalent atomic positions, $\bm{\mu}(\bm{R})$ does not have \rev{periodic} invariance with respect to the periodic boundary conditions.

            In other words, closed integrals do not vanish, but can take up (multiple) polarization quanta, depending on how often the periodic boundaries are crossed.
            We can quantify the change of the  dipole $\Delta \bm{\mu}$ along a closed path $\gamma$ by a line integral. The result of this operation is non-trivial~\cite{rappe}: 
            \begin{align} \label{eq:line-integral}
                \Delta \bm{\mu} \left[\gamma\right] & = \, \oint_{\gamma} d  \bm{\mu}  =
                \oint_{\gamma} d\bm{r} \frac{\partial \bm{\mu}}{\partial \bm{R}} \biggr\rvert_{\bm{R}=\bm{r}} =  e  \bm{l}, 
            \end{align}
            where $\bm{l} = \bm{a} \cdot \bm{n}$ is an element of the Bravais lattice of the system, $ \bm{n} \in\mathbb{Z}^{3}$ and $\bm{a}$ is the matrix of lattice vectors $\bm{a}_{\alpha}$. This result shows that $\Delta \boldsymbol{\mu}=\bm{0}$  when no atom crosses the cell boundaries along $\gamma$. However, if $\gamma$ involves a displacement to a periodic replica at $\bm{R} + \bm{a} \cdot \bm{n}'$ ($\bm{n}'\in\mathbb{Z}^{3}$), the ratio (component-wise) between $\bm{n}$ and $\bm{n}'$ has to be an integer $\mathcal{N}$~\cite{rappe}, i.e.\ 
            \begin{align}\label{eq:oxn-integer}
                \bm{n} = \mathcal{N} \bm{n}'.
            \end{align}
            This relation has been extensively discussed in by Jiang, Levchenko and Rappe in Ref.~\cite{rappe}, where the authors related $\mathcal{N}$ to the atomic oxidation numbers. 
            $\mathcal{N}$  does not depend on the path $\gamma$ as long as no metal-insulator transition occurs and it is defined for each atom of the system. Therefore, $\Delta \boldsymbol{\mu}$ can be non-zero if $\mathcal{N} \neq 0$. This relation also plays a fundamental role for ionic transport~\cite{FrenchRedmer2011}, where  topological arguments relate transport to formal oxidation numbers~\cite{GrasselliBaroni2019}.

            Obtaining $\mathcal{N}$ for different atomic species from a \gls{DFT} calculation is straightforward. One computes the dipole $\boldsymbol{\mu}$ along a path $\gamma$ where the $I^{th}$ atom is displaced from its position $\bm{R}_I$ to $\bm{R}_I + \bm{a} \cdot \bm{n}'$ and evaluates $
                \mathcal{N} =  \Delta \boldsymbol{\mu} \cdot \bm{l} / (e \left|\bm{l}\right|^2)$, 
                with $\bm{l}= \bm{a} \cdot \bm{n}'$.
            In Fig.~\ref{fig:oxn}b and Fig.~\ref{fig:oxn}c we show this calculation graphically for a water molecule in a box and the ferroelectric solid LiNbO$_3$, respectively.
            By defining
            \begin{equation}
            \mathcal{N}(s) =  \frac{ \bm{\mu}(s) \cdot \bm{l} }{ e N_{\rm at} |\bm{l}|^2  }\label{eq:oxn-compute}
            \end{equation}
            we can describe the more general case case where $N_{\rm at}$ atoms of the same oxidation state are  displaced along the same direction.  
            $\mathcal{N}_I$ is the difference between the values of $\mathcal{N}(s)$ upon crossing the periodic boundary.
            The values obtained for $\mathcal{N}_I$ are reported in Fig.~\ref{fig:oxn}b and \rev{Fig.~\ref{fig:oxn}c}, and follow the expected oxidation numbers in these systems.

 
            The previous considerations lead to practical consequences when building a ML model for $\bm{\mu}(\bm{R})$ of a periodic system. 
            Currently, \glspl{ENN} yield the best accuracy to machine-learn properties of materials~\cite{matbench-arxiv}.
            Most commonly, \glspl{ENN} take as inputs the tuple $\left( Z; \bm{R} , \bm{a}_{\alpha} \right)$, where $Z$ is the atomic number, and model the target quantity as a function of the \rev{relative atomic positions} $\Delta\bm{R}_{IJ}\rev{=\bm{R}_{I}-\bm{R}_{J}}$, which are evaluated within the minimum image convention. 
            \rev{This makes any target quantity a function of what we call the \textit{atomic environment} $\boldsymbol{\mathcal{A}}$. Any target quantity which is a function of $\boldsymbol{\mathcal{A}}$ is invariant with respect to any displacement of atom to periodically equivalent sites.}
            
            Going back to Fig.~\ref{fig:oxn}a, because such models require that the function to be learned is differentiable, the discontinuous curve which is obtained by matching $\bm{\mu}$ to assume the same value at equivalent $\boldsymbol{\mathcal{A}}$ cannot be learned in its full domain. If one is only interested in small atomic displacements close to equilibrium, a usual ML model will provide accurate results for $\bm{\mu}$ in this restricted space. However, as one moves into $\boldsymbol{\mathcal{A}}(\bm{R})$ that are far from equilibrium, the ML model will start to change its slope, in order to obey $\bm{\mu}(\boldsymbol{\mathcal{A}}(\bm{R}+\bm{l})) = \bm{\mu}(\boldsymbol{\mathcal{A}}(\bm{R}))$. In an autodifferentiable model, this results in  \rev{\gls{BEC}} $\bm{Z}^*$ \rev{(see Eqs.~\eqref{eq:bec-definition} in Section~\ref{sec:theory})} and consequently forces that start to deviate from the correct value. If more data including environments far from equilibrium are added to the training set, the model will be unable to learn (see \rev{Supplementary Section~S1} for a practical example).

            On the other hand, as one can also see in Fig.~\ref{fig:oxn}a, the multi-valued dipole, where $\bm{\mu}(\boldsymbol{\mathcal{A}}(\bm{R}+\bm{l})) \neq \bm{\mu}(\boldsymbol{\mathcal{A}}(\bm{R}))$, is a continuous smooth function. Learning this function, however, requires a modification to the usual architectures, to be able to describe this multi-valued nature of $\bm{\mu}$.
            In this work, we have slightly modified the \texttt{MACE} equivariant message-passing neural network~\cite{Batatia2022mace,Batatia2022Design} to define a model $\tilde{\bm{\mu}}^\text{MV}$ which depends directly on $\bm{R}$ and $\boldsymbol{\mathcal{A}}$ as follows,
            \begin{align}\label{eq:oxn-model}
                \tilde{\bm{\mu}}^\text{MV}(\bm{R}) = \tilde{\bm{\mu}}(\boldsymbol{\mathcal{A}}) +  e  \sum_I \mathcal{N}_I \bm{R}_I,
            \end{align}
            where $I$ runs over all atoms of the system and $\tilde{\bm{\mu}}$ can be handled by the usual learning procedure, as it depends only on $\bm{\mathcal{A}}$.
            From the definition of $\tilde{\bm{\mu}}^\text{MV}$ and $\bm{Z}^{*}$, we arrive at
            \begin{align}\label{eq:bec-oxn}
                \bm{Z}^{*}_I = \frac{1}{ e } \frac{\partial \tilde{\bm{\mu}}^\text{MV}}{\partial \bm{R}_I} = \frac{1}{ e } \frac{\partial \tilde{\bm{\mu}}}{\partial \bm{R}_I} + \mathcal{N}_I \cdot \mathbbold{1}_{3\times3},
            \end{align}
            In the implementation we used in this work, we modified the output of the forward pass of \texttt{MACE} to include the second term on the right of Eq.~\eqref{eq:oxn-model}. In this way, we always calculate $\tilde{\bm{\mu}}^\text{MV}$ in the training and use that also in the calculation of the loss-function. Training is successful and accurate even for datasets containing very out of equilibrium structures of H$_2$O and LiNbO$_3$, and $\bm{Z}^*$ obtained by autodifferentiation is also accurate (see Supplementary Section~\rev{S1} and Section~\ref{sec:methods}). 

            The successes of previous ML models of dipoles and \gls{BEC} are numerous~\cite{Gigli2022, falletta2024unified, Jana2024,10.1063/1.5141950,PhysRevLett.120.036002,PhysRevB.102.041121, Unke2019,VeitCeriottiJCP,gastegger2017machine, schmiedmayer2024derivative, Schienbein2024}. They are successful because they either treat diffusive systems where the combined $\mathcal{N}$ (summing over all diffusing atoms) is zero, or address problems where the displacements along modes \rev{that result in significant variations of the dipole } are not large and the system stays close to equilibrium along that coordinate.
            We note that models targeting to learn $\bm{Z}^{*}$ can also be problematic if they miss the last term in Eq.~\eqref{eq:bec-oxn} and they are not guaranteed to obey the acoustic sum rule for this quantity\rev{~\cite{ASR}}.
            The approach we present encompasses and extends these models to be generally applicable. This is for instance important for ionic diffusion~\cite{GrasselliBaroni2019} and strongly non-equilibrium phonon driving of ferroelectric materials, as we show below.


        \subsection{Dielectric Properties and Electrofreezing of Water}

            \begin{figure*}[!htb]
                \centering
                \includegraphics[width=0.95\textwidth]{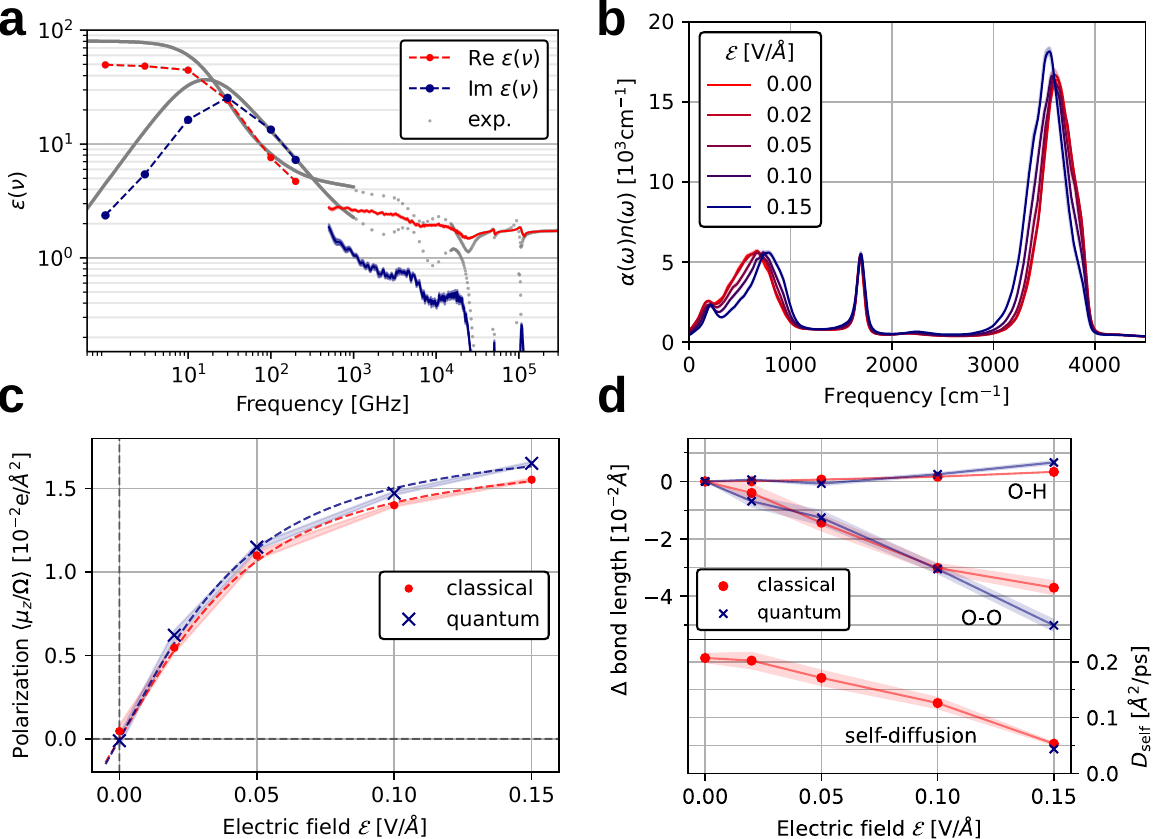}

                \caption{Liquid water properties from ML-based molecular dynamics simulations under applied electric fields. 
                \textbf{(a)} Calculated real and imaginary parts of the dielectric permittivity at 300~K in the 1-135000~GHz frequency range (see text). The gray points correspond to experimental data at 20$^{\circ}$C from Ref.~\cite{watericebook}. Dashed lines serve as guides to the eye.
                \textbf{(b)} IR absorption spectra of liquid water at 300~K and varying electric-field strengths. %
                \textbf{(c)} Average of the polarization along the direction of applied electric fields of varying strengths obtained from simulations with classical nuclei (red) and quantum nuclei (blue). Dashed lines represent fits assuming a model of classical non-interacting dipoles.
                \textbf{(d)} \rev{Top:} OH bond length and O-O nearest neighbor distance with varying field strengths in simulations with classical (red) and quantum (blue) nuclei. \rev{Bottom:} Self-diffusion coefficient of water calculated from the root-mean-square displacements of oxygen atoms at varying field strengths. Shaded areas in all panels correspond to the standard error.
                }
                \label{fig:water-dielectri-freez}
            \end{figure*}

            We first apply the ML models we developed to study liquid water at room temperature and standard density. Water is of fundamental importance to life and, as such, it has been addressed numerous times in literature~\cite{water_ceriotti_2016,watericebook,Eltonpccp2017}, with many peculiar features related to its dielectric properties and vibrational dynamics under applied electric fields still not fully resolved. The polarization of water results from a complex interplay between changes in the dipoles of  individual water molecules and the correlated dynamics of multiple molecules in the liquid state. The results we present below were obtained by coupling the ML model we developed for the dipoles with a \gls{mlip} model for the energies and forces, see  Section~\ref{sec:methods}.

            Our implementation allows nanosecond long simulations and the application of dynamic and static electric fields, going well beyond the time-scales reached by existing \textit{ab initio} techniques. It is thus possible to efficiently simulate the frequency-dependent dielectric function $\varepsilon(\nu)$ of liquid water, as shown in Fig.~\ref{fig:water-dielectri-freez}a, in the range of $0.001-135$ THz ($0.03-4500$ cm$^{-1}$). Spanning this large frequency range was made even more efficient through the combination of two simulation techniques that are allowed by our methodology. Above 100 GHz, the real and imaginary parts of $\varepsilon(\nu)$  were calculated through the fluctuation-dissipation relation~\cite{susEexplicit,PhysRevB.77.012102,Gigli2022} (see Supplementary Section~\rev{S2}), while the region below 100 GHz was obtained by a more efficient non-equilibrium simulation technique~\cite{susEexplicit} (see Section~\ref{sec:methods}). We could then simulate the onset of the Debye relaxation, which reduces the value of $\varepsilon(\nu)$ at frequencies larger than 10 GHz~\cite{watericebook}. The results shown in Fig.~\ref{fig:water-dielectri-freez}a
            agree quite well with experimental data~\cite{watericebook} shown in the same figure, despite a slight underestimation of the static dielectric constant, which we attribute to the training data included in the $\tilde{\bm{\mu}}^\text{MV}$ model in this case. As we show in Supplementary Section~\rev{S3}, adding dipoles computed in periodic bulk-water structures to the training dataset shifts the computed results closer to the experimental data and better reproduces previous \textit{ab initio} results~\cite{MarxDielectric2021}. Besides these differences, the results substantiate that our approach is able to qualitatively capture the atomistic mechanisms of the Debye relaxation in water and should be a useful tool to settle the debate about the spatial extent of dynamical molecular correlation involved in this process, without having to resort to empirical potentials for large system sizes~\cite{Eltonpccp2017,MarxDielectric2021}.

            In Fig.~\ref{fig:water-dielectri-freez}b we show the IR spectra of water under applied static electric fields, as obtained from the Fourier transform of dipole-autocorrelation time-series, simulated with molecular dynamics. The IR spectrum is closely related to ${\rm Im} \, \varepsilon(\nu)$ shown in Fig.~\ref{fig:water-dielectri-freez}a.
            The spectra in Fig.~\ref{fig:water-dielectri-freez}a are in excellent agreement with the ones recently reported in Ref.~\cite{Schienbein2024}, which employed a similar \textit{ansatz} of linear-coupling to the electric-field as we do here.

            With increasing field strength, there is a pronounced blue-shift of the libration band below $1000$ cm$^{-1}$, a very small blue-shift of the bending band at $\approx1700$ cm$^{-1}$ and of the combination band at $\approx2200$ cm$^{-1}$, and a pronounced red-shift  of the OH stretching band at $\approx3600$ cm$^{-1}$, effectively narrowing the spectral range.
            \rev{The frequency shifts of these bands at the field strength of 0.15 V/\AA~ amount approximately to $113$, $7$, and $-75$~cm$^{-1}$, respectively.}
            Such shifts are common features of the vibrational Stark effect in anharmonic systems~\cite{BoxerStark1995}. 
            The red-shift of the OH stretching band is due to a strengthening of H-bonds, evidenced by shorter O--O distances, as shown in Fig.~\ref{fig:water-dielectri-freez}d. The blue-shift of the libration band is due to the breaking of molecular rotational isotropy at increasing field strengths, as also discussed in Refs.~\cite{Schienbein2024,Cassone2024,FuteraEnglish2017}. 
            
            Related to this blue-shift in the libration band, while the band we calculate is in excellent agreement with the one shown in Ref.~\cite{Schienbein2024} at all field strengths, both results do not contain the new peak just below $1000$ cm$^{-1}$ at higher electric field strengths, reported by Cassone and coworkers and by Futera and English~\cite{Cassone2024, FuteraEnglish2017}. This band is ascribed to a libration motion that is enhanced because of the preferential orientation of water along the electric field direction.
           As we also observe the preferential orientation of water molecules in our study (Fig.~\ref{fig:water-dielectri-freez}c), these differences must depend on the underlying methodology. We propose that the dynamics involved in the enhancement of this peak may be strongly dependent on the different underlying exchange-correlation functionals or on the inclusion of second-order electric-field coupling in the nuclear Hamiltonian, which are not explicitly considered in our work or in that of Ref.~\cite{Schienbein2024}. Because it is straightforward to augment our methodology to account for these terms, these considerations will be the subject of future work.

           We now analyze the impact of nuclear quantum effects (NQE) together with the application of electric fields in water. To the best of our knowledge, the importance of these effects at higher electric field strengths has not been previously discussed. 
           \rev{The narrowing of the IR spectrum with increasing field strength is more pronounced when accounting for NQE, as shown in Supplementary Section~S4. The frequency shifts at 0.15 V/\AA~field strength amount approximately to $135$~cm$^{-1}$ for the libration band and $-175$~cm$^{-1}$ for the stretching band. It is interesting to look closer at the origin of this enhanced effect.}
           In Fig.~\ref{fig:water-dielectri-freez}c we report the average polarization of liquid water along the field direction and at varying field strengths $\mathcal{E}$, from classical-nuclei and \rev{\gls{PIMD}}. While at zero-field there is no net polarization, the interaction with the field causes a net polarization to appear along the field direction. 
            The increase of the average polarization for $\mathcal{E}\gtrsim 0.04$ V/\AA~ is non-linear, in agreement with Ref.~\cite{Schienbein2024}. This behavior can be understood by analyzing the thermodynamics of an ensemble of classical non-interacting dipoles in an external field.
            The analytical expression of the partition function of this system can be found in textbooks~\cite{rigamonti2007structure}, and the total polarization of the system is expressed as a Langevin function that depends on $\mathcal{E}$ and the molecular dipoles. By fitting to this model (dashed lines in Fig.~\ref{fig:water-dielectri-freez}c), we conclude that NQE make water molecules easier to polarize upon electric field application. The saturation value of the average polarization is higher when including NQE, because the enthalpic term of interaction with the field can better compensate the thermal entropy of the liquid. 

            The impact of NQE on the average polarization correlates with the changes in structural properties of water, shown in Fig.~\ref{fig:water-dielectri-freez}d. With increasing $\mathcal{E}$ the OH bond length increases while the O-O nearest-neighbor distances decrease, consistent with the strengthening of the H-bonds. These effects are much more pronounced at $\mathcal{E} = 0.15$ V/\AA~ when including NQE. The stronger electric fields thus shift the balance of the competing NQE in water~\cite{Habershon_2009}, causing NQE to work towards considerably strengthening the H-bond interaction. 
            
            We also calculated the water self-diffusion \rev{coefficient $D_{\mathrm{self}}$} at different $\mathcal{E}$, shown in Fig.~\ref{fig:water-dielectri-freez}d. Consistent with the phenomenon of electrofreezing~\cite{Cassone2024}, we observe this coefficient to decrease from around \rev{0.20} \AA$^2$/ps at zero-field to \rev{0.05} \AA$^2$/ps at $\mathcal{E} = 0.15$ V/\AA\,\rev{in the classical case. Consistent with} the previous considerations, NQE also enhance this effect, causing a faster decrease in \rev{$D_{\mathrm{self}}$} with increasing field strength. \rev{At zero-field, previous reports have shown that NQE increase $D_{\mathrm{self}}$ with respect to the classical simulations by about 12\%~\cite{Rossi2014}, and here we show that it is instead smaller than the classical value at $\mathcal{E} = 0.15$ V/\AA, as shown in Fig.~\ref{fig:water-dielectri-freez}d}.

        \subsection{Phase Transition and Non-equilibrium Phonon Driving in LiNbO$_3$}\label{subec:PT}

            \begin{figure*}[!htb]
                \centering
                \includegraphics[width=\linewidth]{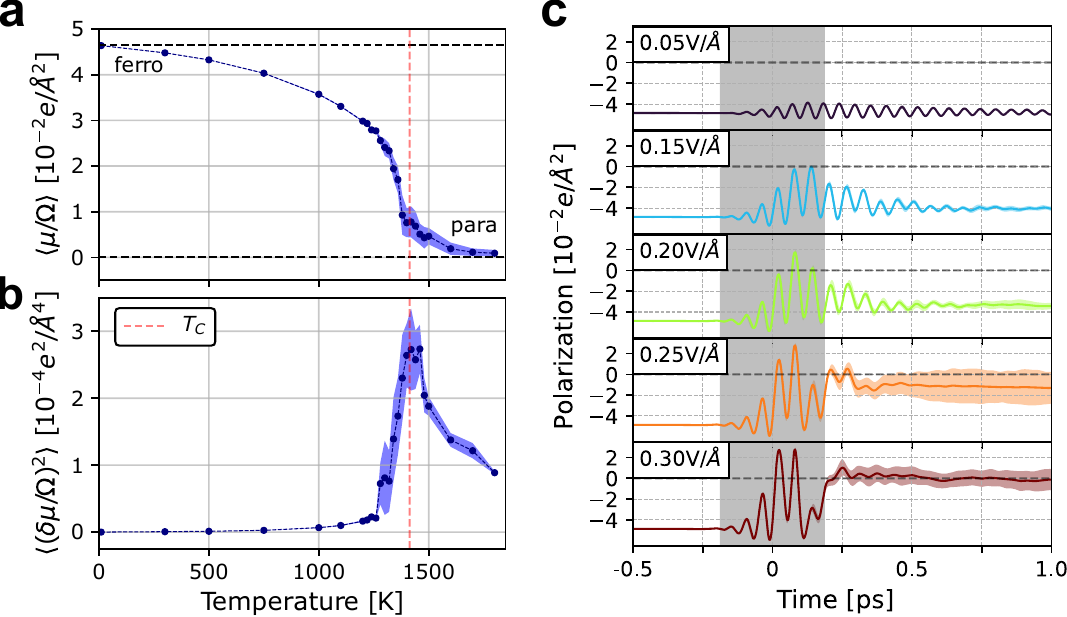}
                \caption{Phase transition and non-equilibrium properties of LiNbO$_3$ from ML-based molecular dynamics simulations.
                \textbf{(a)} Thermal average of the dipole at varying temperature. The polarization shows a transition from a ferroelectric state (at low temperature) to a paraelectric state with a Curie temperature (red dashed line) close to the experimental value $T_C=1413$~K~\cite{Smolenskii}.
                The polarization was computed as the dipole $\bm{\mu}$ per unit volume $\Omega$ along the direction $\left[111\right]$.
                \textbf{(b)} The polarization fluctuations $\delta \mu/\Omega$ shows a broadened peak around the Curie temperature as expected from a second-order phase transition simulated in a finite-system. Shaded areas correspond to the standard error over uncorrelated trajectories.
                \textbf{(c)} Ultrafast dynamics of the polarization of LiNbO$_3$ under vibrational excitation with a monochromatic THz laser pulse with $\nu$ = 18~THz and FWHM = 188~fs and at varying maximum intensities. The $y$ axis shows the polarization with respect to the paraelectric state. The gray area represents the window of time during which the pulse is applied and the origin of time is at the center of the pulse.
                }
                \label{fig:LiNbO-phase-transition}
            \end{figure*}

            Next, we show the generality of the ML approach developed here to address the ferroelectric to paraelectric phase-transition in solid LiNbO$_3$, a widely employed  ferroelectric perovskite with a Curie temperature of $T_C = 1413$~K~\cite{LNCurieTemp1,LNCurieTemp2,Smolenskii}. The study of this system presents a  counterpart to the previous results on water, as polarization changes are predominantly induced by the large-amplitude motion of the Li$^+$ ions. In this system, a single-valued model for $\bm{\mu}$ would readily fail. As we show in the Supplementary Section~\rev{S1}, when attempting to train a single-valued model on the dataset of LiNbO$_3$ including structures where the Li$^+$ atoms have been driven far from their equilibrium positions, the training is simply not successful.

            We have evaluated the mean and the variance (fluctuations) of the dipole in the temperature range of $10$ to $1800$~K. The results are shown in  Fig.~\ref{fig:LiNbO-phase-transition}a and \rev{Fig.~\ref{fig:LiNbO-phase-transition}}b.
            Both plots show the typical behavior of a second-order phase transition~\cite{Smolenskii}. The average value of the order parameter (the dipole) decreases to zero above the Curie temperature (Fig.~\ref{fig:LiNbO-phase-transition}a), while its fluctuations show a sharp peak at the same temperature (Fig.~\ref{fig:LiNbO-phase-transition}b). Finite size effects cause this peak to broaden with respect to the expected $\delta$-function. 
            The analysis of the atomic displacements presented in Supplementary Section~\rev{S5}, which are in agreement with simulations reported in Ref.~\cite{LNLT}, confirm that the simulations indeed capture the ferroelectric to paraelectric phase-transition.

            The  phase-transition temperature inferred from Fig.~\ref{fig:LiNbO-phase-transition}a and \rev{Fig.~\ref{fig:LiNbO-phase-transition}}b matches quite well the experimental Curie temperature $T_C$, which is likely to be partially fortuitous, because our simulations do not include the effects of thermal expansion. These effects are expected to increase $T_C$ by around 100 K~\cite{LNLT}. Nevertheless, the reasons why the value we calculate is so close to the experimental result are that we properly account for all orders of anharmonic coupling between vibrational modes and that we accurately model its polarization changes, including its multi-valued nature. The absence of anharmonic couplings was previously shown to cause an underestimation of 250-300 K for $T_C$~\cite{LNLT,LNLT2}. 
       
            LiNbO$_3$ has been at the center of active scientific debates due to the possibility of reversing its polarization using THz laser pulses which induce a non-thermal excitation of an IR-active mode $Q_{\rm IR}$ that strongly couples to a polarization-reversal mode $Q_{\rm P}$, as theoretically proposed in Ref.~\cite{subedi2015} and partially realized in Ref.~\cite{MankowskiCavalleri2017}.
            To date, theoretical studies modeling this effect~\cite{PhysRevLett.123.129701,Chen2022,vonHoegen2018} 
                were based on approximations to the \gls{PES} which consider two or only a few phonon modes and the lowest-order anharmonic couplings between them.
                We have employed the ML models developed in this work to address this question, without relying on such approximations. 
                
                We have performed simulations where we applied a time-dependent monochromatic pulse with frequency $\nu=18$~THz enveloped with a Gaussian function with  FWHM=188~fs and different maximum field intensities $\mathcal{E}_{\mathrm{max}}$ in the range $0.05 - 0.30$~V/\AA. These values were chosen based on the reported experimental setups~\cite{MankowskiCavalleri2017,vonHoegen2018} (see also Supplementary Section~\rev{S6}). The excitation is in close resonance with the 18.4 THz $Q_{\rm IR}$ mode, while the proposed $Q_{\rm P}$ mode lies at $7.4$ THz. These modes are depicted in Supplementary Section~\rev{S7}. 

                As shown in the Supplementary Sections~\rev{S7} and \rev{S8}, we observe the non-linear coupling of the $Q_{\rm IR}$ mode that is directly excited by the pulse at $\nu=18$~THz with the $Q_{\rm P}$ mode, which starts accumulating energy with a delay of some hundreds of fs. In addition we observe that only two other $\Gamma$-point modes, of A$_1$ symmetry, at 8.2 and 10.0 THz get excited at later times. These modes are also IR active and, in particular the mode at 8.2 THz contributes to the polarization-reversal, attesting to the complex dynamical coupling in this system. In Fig.~\ref{fig:LiNbO-phase-transition}c we show the time-dependent variation of the polarization at times previous to, during (gray shaded area) and after the laser pulse is applied.
                The value of the polarization deviates from the thermal equilibrium value as soon the laser pulse starts acting on the system.
                We can recognize three different regimes, depending on the maximum intensity of the laser pulse.
                (i) At low intensities such as $0.05$~V/\AA, the polarization is only slightly perturbed and the system remains ferroelectric.
                (ii) At higher intensities ($0.15$ and $0.20$~V/\AA), the system is transiently driven to the paraelectric state (with zero polarization) during the laser pulse, but returns to the original ferroelectric state with a relaxation time of about $300$~fs, during which coherent dynamics can still be observed. 
                (iii) At extremely high fields ($0.25$ and $0.30$~V/\AA), the system transiently reaches higher values of reversed polarization but never fully switches, and after the pulse is switched off it transitions to the paraelectric state with incoherent dynamics because the amount of energy absorbed effectively heats the system above $T_C$. 

                We do not observe a full coherent switching of the polarization in LiNbO$_3$.
                This is in agreement with experiments~\cite{MankowskiCavalleri2017}, which could not fully achieve a polarization reversal with this type of phonon driving. This has been attributed to the  presence of depolarization fields~\cite{depolarizationLINOBATE}, which prevent the full reversal, and which we can capture in the large unit cells used for the dynamics in this study.
                However, Fig.~\ref{fig:LiNbO-phase-transition}c shows that, within the pulse duration and at intense enough fields, it is possible to drive the system coherently to the paraelectric state.
                We propose that this is likely the explanation for the dip in the second-harmonic generation signal during the application of the field observed in the experiments reported in Ref.~\cite{MankowskiCavalleri2017}.


        \section{Discussion}\label{sec:discussion}

            The results we have shown attest to the reliability and efficiency of the dipole ML model we developed. Together with a molecular dynamics protocol that captures the coupling of time-dependent and static electric fields with matter, it can deliver accurate properties and new physical insights for liquids and solids, both at equilibrium and driven far from equilibrium. This formulation thus extends the applicability realm of several ML models previously proposed, being applicable to isolated and periodic systems on the same footing and automatically satisfying charge conservation. 

            Regarding the specific applications we have shown, we would like to highlight a few points. In water, while the changes in the polarization behaviour due to NQE under applied electric fields are seemingly small, the magnitude of such changes can lead to vastly different state-points of phase transitions. We expect these effects to be particularly significant for superionic phases of water, which can be tuned by electric fields and by nanoconfinement~\cite{ravindra2024,FuteraSciAdv2020}. A study of the stability and dynamics of superionic phases under applied electric fields with ML methods would require a multi-valued dipole model such as the one developed here.
            
            Indeed, we have shown that the necessity of employing a multi-valued ML model is paramount to capture phase transitions and general metastable and non-equilibrium states in LiNbO$_3$. With an extension of our method to account for the coupling of electric fields with the lattice degrees of freedom, we also expect to be able to capture non-equilibrium dynamics of piezoelectric solids, considerably increasing the breadth of simulation tools in the area of vibrational ultrafast driving of materials.
            
            In summary we expect that the simple technique we proposed, which incorporates the fundamental topological properties of the polarization of periodic systems into ML models, will bring significant advantages to the simulation of a range of liquids and solids that vastly surpass the examples shown here.

    \section{Methods \label{sec:methods}}

    \subsection{\rev{Theory of Molecular Dynamics Driven by Electric Fields \label{sec:theory}}}

        When the applied electric field $\bm{\mathcal{E}}$ is small compared to the internal fields, one can apply the \gls{EDA} within the linear-response regime. In this case, the nuclear Hamiltonian can be expressed as~\cite{Stengel2009,PhysRevLett.89.117602, UmariPasquarello2002}
        \begin{align}
            \mathcal{H}\left(\bm{p},\bm{R}\right) = T\left(\bm{p}\right) + V\left( \bm{R} \right) - \boldsymbol{\mathcal{E}} \cdot \bm{\mu} \left( \bm{R} \right),\label{eq:ham}
        \end{align}
        where $T(\bm{p})$ is the kinetic energy, $\bm{p}$ the nuclear momenta, $\bm{R}$ the nuclear coordinates,  $V(\bm{R})\equiv E[n_0]$ is the \gls{BO} \gls{PES} obtained from the ground-state electronic density $n_0$, and $\bm{\mu}(\bm{R}) \equiv \bm{\mu}[n_0]$ is the system dipole.

        The forces on the nuclei depend on the external field $\boldsymbol{\mathcal{E}}$ through 
        \begin{align}
            \bm{F}
            = \, - \frac{\partial \mathcal{H}}{\partial \bm{R}}
            = - \frac{\partial V}{\partial \bm{R}} + \boldsymbol{\mathcal{E}} \cdot \frac{\partial \bm{\mu}}{\partial \bm{R}}\label{eq:forces}.
        \end{align}
        The first term on the right-hand side is the usual \gls{BO} force. The second term can be expressed in terms of the atomic-polar tensors~\cite{Schienbein2024,Kapil2024}, also called  \gls{BEC} $\bm{Z}^{*}$, 
        \begin{align}\label{eq:bec-definition}
            \bm{Z}^{*}
            = \, \frac{1}{e}\frac{\partial \bm{\mu}}{\partial \bm{R}} = \frac{1}{e}\frac{\partial \bm{F}}{\partial \boldsymbol{\mathcal{E}}}
        \end{align}
        where $e$ is the elementary charge. In this formulation, if the electric field $\boldsymbol{\mathcal{E}}(t)$ varies in time, the Hamiltonian simply gains a dependence on time $\mathcal{H}(\bm{p},\bm{R},t)$. Molecular dynamics proceeds through the calculation of the \gls{BO} forces and the \gls{BEC} at each time-step for the numerical integration of the equations of motion. 
        
        For a given atomic configuration, the \gls{BEC} can be computed from any of the two 
        expressions in Eq.~\eqref{eq:bec-definition}.
        The (relative) efficiency of these approaches strongly depends on the system, on the chosen electronic-structure code, and on details of the implementations. Evidently, the expressions can be calculated by finite differences or by perturbation theory~\cite{baroni2001phonons,PhysRevB.55.10355}. While the latter is elegant and can be quite efficient especially for systems where symmetry can be exploited, the former ensures a natural handling of electronic-structure level of approximation, such as different density-functionals, relativistic corrections and other terms that are cumbersome to evaluate analytically. In any event, the calculation of $\bm{Z}^{*}$ always represents an added cost on the already costly electronic-structure calculation \rev{of the dipole}.

        We circumvent the expensive first-principles calculation of $\bm{Z}^{*}$ that enter Eq.~\eqref{eq:forces} by developing a ML model that is trained only on $\bm{\mu}$ and that is autodifferentiable~\cite{autodiff} with respect to its input quantities, the nuclear coordinates $\bm{R}$. Training on $\bm{\mu}$ presents some advantages, including the ease to produce data for this quantity with diverse electronic-structure methods and the guarantee that the acoustic sum rule~\cite{ASR} for $\bm{Z}^{*}$ in the model is preserved, making it strictly translationally invariant.

        The evolution of the nuclear equations of motion including time-dependent and independent applied electric fields was obtained by an implementation developed by us in the \texttt{i-PI} code, which was recently described in Ref.~\cite{iPI2024}. We refer the reader to that publication for technical details. All molecular-dynamics simulations were run with \texttt{i-PI} as the driving code communicating with the \gls{ML} models, which provided all components of the force.

        \begin{table}[t]
            \centering
             \begin{tabular}{rrr}
                \toprule
                parameters & H$_2$O & LiNbO$_3$\\
                \midrule
                n. atoms & 3 & 10 \\
                $a=b=c$ [\AA] & 4 & 5.48  \\
                $\alpha=\beta=\gamma$ [deg] &  90 &  55.91 \\
                \bottomrule
            \end{tabular}
            \caption{Unit cell parameters used for the calculation of the oxidation numbers as shown in Section~\ref{sec:dipoles}.}
            \label{table:oxn-cell}
        \end{table}
        
        \begin{table}[t]
            \centering
            \begin{tabular}{cccc}
                \toprule
                species & direction & n. atoms & $\mathcal{N}$ \\
                \midrule
                \multicolumn{4}{c}{H$_2$O} \\
                \hline
                H  & $\left[0,\phantom{-}1,\phantom{-}0\right]$ & 2 & \phantom{-}1 \\
                O  & $\left[0,\phantom{-}1,\phantom{-}0\right]$ & 1 & -2 \\
                \hline
                \multicolumn{4}{c}{ LiNbO$_3$} \\
                \hline
                Li & $\left[1,-1,\phantom{-}0\right]$ & 2 & \phantom{-}1 \\
                Nb & $\left[1,-1,\phantom{-}0\right]$ & 2 & \phantom{-}5 \\
                O  & $\left[1,\phantom{-}1,\phantom{-}1\right]$ & 6 & -2 \\
                \bottomrule
            \end{tabular}
            \caption{Computational details used to evaluate the oxidation numbers (as presented in Section~\ref{sec:dipoles}) of all the atomic species in water and LiNbO$_3$. The table reports the direction, on the basis of the lattice vectors, along which the atoms were displaced, the total number of atoms of that species, and the evaluated oxidation numbers.}
            \label{table:oxn}
        \end{table}

        \begin{table}[t]
            \centering
            \begin{tabular}{clrrc}
                \toprule
                 quantity & unit & H$_2$O & LiNbO$_3$ & \\
                \midrule
                energy & meV/atom & 1.10*  & 0.85 &\\
                \hline
                forces & meV/\AA  & 31.19* &  27.51 &\\
                \hline
                dipole & mD/atom  & 3.12\phantom{*}\,   & 11.38 &\\
                \bottomrule
            \end{tabular}
            \caption{Performance of the trained models for liquid water and LiNbO$_3$. The values marked with an asterisk represent RMSE on the training dataset, all other values represent RMSE on the test set. See Supplementary Section~\rev{S9} for further details and information.}
            \label{table:accuracy}
        \end{table}   

        \subsection{\rev{Dipoles in Aperiodic and Periodic Systems \label{sec:dipoles}}}

        While the calculation of the nuclear contribution $\boldsymbol{\mu}_n$ to $\bm{\mu} = \boldsymbol{\mu}_e + \boldsymbol{\mu}_n$ presents no issues, the electronic contribution $\boldsymbol{\mu}_e$ can be more cumbersome to calculate. In aperiodic systems, $\boldsymbol{\mu}_e$ can be obtained through a real-space integral involving the ground-state electronic density $n_0(\bm{r})$
            \begin{align}\label{eq:dipole-e-no-pbc}
                \boldsymbol{\mu}_e = - e \int_{\mathbb{R}^3} d\bm{r} \, n_0(\bm{r}) \bm{r}.
            \end{align}
    
            In periodic, condensed-phase systems, $\boldsymbol{\mu}_e$ cannot be directly evaluated by the expression above, as restricting the integration domain to the primitive unit cell would yield boundary-sensitive results. 
            The dipole is related to the polarization $\bm{P}$ of a periodic system, in that the polarization is the dipole per unit volume, $\bm{P}=\boldsymbol{\mu}/\Omega$. The modern theory of polarization (MTP)~\cite{SPALDIN20122,PhysRevB.47.1651,Resta2007} provides a proper definition of this quantity in periodic systems, showing that the electronic contributions to $\bm{P}$ \rev{are related to a} geometric phase of the wavefunction. There are several different (but equivalent) ways to evaluate $\boldsymbol{\mu}_e$ within MTP, which are all related to a way of evaluating the Berry \rev{phase}~\cite{PhysRevB.47.1651, Resta2007, PhysRevB.56.12847}.
            As a consequence, the components of $\bm{\mu}$ corresponding to each lattice vector $\bm{a}_{\alpha}$ can be defined only up to quantum of polarization $e \left|\bm{a}_{\alpha}\right|$ and the dipole along $\bm{a}_{\alpha}$ is 
            \begin{equation} \label{eq:dipole-quantum}
            \mu_{\alpha} \equiv \mu_{\alpha} \mod e|\bm{a}_\alpha|. 
            \end{equation}

        \subsection{Water \rev{Simulation Details}}
        
            We have evaluated the oxidation numbers of the H and O atoms of a single water molecule following Section~\ref{sec:dipoles}.
            A cubic simulation box of $4$~\AA~ (whose parameters are reported in Table~\ref{table:oxn-cell}) was used. Figure~\ref{fig:oxn}b was obtained considering $100$ configurations, where the H atoms were uniformly displaced along the direction reported in Table~\ref{table:oxn} from the relaxed initial configuration, while the O atom was fixed, and vice-versa.
            The \gls{DFT} calculations were performed with the revPBE functional, no spin polarization, a $3\!\times\!3\!\times\!3$ $\bm{k}$-grid, atomic ZORA relativistic treatment~\cite{10.10631.467943}, and the \texttt{intermediate} basis set in the \texttt{FHI-aims}~\cite{fhiaims2009} code. %
            The polarization components were calculated~\cite{carbogno2025polarisationborneffectivecharges} on a denser $\bm{k}$-grid, namely $40\!\times\!10\!\times\!10$ for the component along $\bm{a}_1$, $10\!\times\!40\!\times\!10$ for $\bm{a}_2$, and $10\!\times\!10\!\times\!40$ for $\bm{a}_3$.

            The \gls{mlip} for liquid water was trained with a \texttt{MACE}~\cite{Batatia2022mace,Batatia2022Design} model on the data provided in Ref~\cite{Cheng2018AbIT}, i.e. energy and forces computed using the revPBE0 functional with Grimme D3~\cite{Grimme2010,Grimme2011} dispersion corrections.
            We have trained a \texttt{MACE} \gls{mlip} with a cutoff radius of $6$~\AA, $2$ layers, $64$ embedding channels, \rev{$\ell=3$} as the highest spherical harmonics, a maximum of \rev{$L=0$} for each message, and correlation order $3$ at each interaction layer.

            The dataset for training the dipole model was obtained from 3 \texttt{NVT} simulations (stochastic-velocity-rescaling thermostat SVR ~\cite{svr2007}; $\tau= 100$ fs; $T = 330, 360, 390$~K) with \texttt{i-PI}~\cite{iPI2024} and the trained \gls{mlip} as a force provider. We ran $100$~ps for each run with a time-step of $1$~fs.
            The simulation box was cubic with lattice vectors of length $14.9365$~\AA~ and $111$ molecules, corresponding to a density of $0.9965$~g/m$^{-3}$.
            The \gls{FPS} algorithm, as implemented in the \texttt{librascal} library~\cite{librascal} and based on \texttt{SOAP} descriptors as reported in Ref.~\cite{cersonsky2021improving} was used to select the $1000$ most diverse structures from all  trajectories.

            We used the structures in the unit-cell in an aperiodic setting (effectively water droplets) and computed the dipole using Eq.~\eqref{eq:dipole-e-no-pbc} with the revPBE functional. This was done solely to increase efficiency of data generation.
            We have split the data with a ratio of $80/20$ for training and validation, after having separated 20\% of the full dataset for testing.             
            The dipole model was trained using a modified \texttt{MACE} architecture developed by us and available in Ref.~\cite{eliamacerelease}, with a cutoff radius of $6$~\AA, $2$ layers, $4\!\times\!0e\!+\!4\!\times\!1$o as irreps for the hidden node states, \rev{$\ell=2$} as the highest spherical harmonic, correlation order of $3$ at each layer. We added the oxidation numbers contribution according to Eq.~\eqref{eq:oxn-model} in the training.
            The performance of these models are reported in Table~\ref{table:accuracy}.
            
            The IR absorption spectra of liquid water and field-dependent self-diffusion coefficients (Fig.~\ref{fig:water-dielectri-freez}a and d) were obtained by first equilibrating the system for $100$~ps at $300$~K (Langevin thermostat,  $\tau=50$~fs, $dt = 0.25$~fs) and then applying the constant electric fields following Eqs.~\eqref{eq:ham} and \ref{eq:forces}. The \gls{mlip} together with the ML dipole model were used to obtain the forces. We picked $16$ random snapshots from this trajectory and used them to start 16 \texttt{NVE} trajectories of $40$~ps each. 
            The IR absorption spectra were obtained from the Fourier transform of the autocorrelation of the dipole derivatives~\cite{McQuarrie-Book}. The diffusion coefficient was obtained by computing the \gls{MSD} of the O atoms for each time-step and taking the average value of its time-derivative on the last $10$~ps, when the behavior is already linear with time
            
            The high-frequency part of the dielectric susceptibility was computed with the fluctuation-dissipation relation of the dipole and dipole-derivative equilibrium correlation function~\cite{susEexplicit,PhysRevB.77.012102,Gigli2022}, using the same data as for the infrared absorption spectrum with zero electric field.
            The low-frequency part of the dielectric susceptibility was computed using the non-equilibrium direct electric field method described in Ref.~\cite{susEexplicit}. We simulated \texttt{NVT} trajectories of \rev{1}~ns (SVR thermostat,  $\tau=100$~fs, $dt=0.5$~fs) including a time-dependent monochromatic external electric field of $\mathcal{E}_{\rm max} =0.04$~V/\AA, with frequencies of $1, 3, 10, 30, 100$ and $200$~GHz.
            The time-dependent behavior of the  (induced) polarization at each frequency was fitted with a function $P_{\rm ind}\cos\left(2\pi\nu-\phi\right)$ to obtain the parameters $P_{\rm ind}$ and $\phi$. The introduction of the 3$^{\rm rd}$ and 5$^{\rm th}$ harmonics in the fit, as done in Ref.~\cite{susEexplicit}, resulted in negligible corrections to the fitted values. 
            The real and imaginary parts of the dielectric susceptibility were then  computed as $P_{\rm ind}\cos\left(\phi\right)/\varepsilon_0\mathcal{E}_{\rm max}$ and $P_{\rm ind}\sin\left(\phi\right)/\varepsilon_0\mathcal{E}_{\rm max}$ respectively.
            
            The dipole and structural data in Figs.~\ref{fig:water-dielectri-freez}c and d were derived from $16$ \texttt{NVT} trajectories for each different electric field of $10$~ps each. 
            The static structural properties related to electrofreezing were also calculated including nuclear quantum effects with \gls{PIMD}.
            We ran the same number and length of trajectories for each electric field intensity, for 30~ps, using $8$ ring-polymer replicas, a time-step of 0.25~fs, and the PIGLET thermostat~\cite{piglet}.

        \subsection{LiNbO$_3$ \rev{Simulation Details}}

            The oxidation numbers of the atoms in LiNbO$_3$ presented in Fig.~\ref{fig:oxn}c were computed by using the fully relaxed unit cell whose parameters are reported in  Table~\ref{table:oxn-cell} with space group $R3c$. 
            The atoms were displaced along the directions reported in Table~\ref{table:oxn}.
            We employed the PBEsol functional, a $\bm{k}$-grid of $3 \times 3 \times 3$, no spin polarization, atomic ZORA relativistic treatment, and the \texttt{intermediate} basis set in \texttt{FHI-aims}. All atoms of the same species were displaced together.
                          
            The LiNbO$_3$ dataset for training the \texttt{MACE} \gls{mlip} and dipole models was obtained from $3$ {\it ab initio} \texttt{NVT} trajectories of 20~ps at 500~K, 1000~K, 1500~K (SVR thermostat, $\tau=100$~fs, $dt = 2$~fs) with a simulation box containing $30$ atoms and lattice vectors $a,b,c=5.14, 5.14, 13.82$~\AA~ and angles $\alpha,\beta,\gamma= 90^\circ, 90^\circ, 120^\circ$.
            We used the \gls{FPS} algorithm to select $5000$ structures.
            For these structures, we computed the polarization using the Berry phase formalism implemented in~\texttt{FHI-aims}~\cite{carbogno2025polarisationborneffectivecharges}.
            The values of the dipoles have been branch-matched as described in  Supplementary Section~\rev{S10}.
            
            The \gls{mlip} consists of a \texttt{MACE} model with a cutoff radius of $6$~\AA, $2$ layers, \rev{$\ell=2$} as the highest spherical harmonics, a maximum of \rev{$L=0$} for each message, correlation order $3$ at each layer. The parameters for the \texttt{MACE} dipole model were the same as the ones used for water, but adjusting for the  oxidation numbers of LiNbO$_3$.
            The final errors on energies, forces and dipoles for these models are reported in Table~\ref{table:accuracy}.

            The simulations regarding the  ferroelectric to paraelectric phase transition  in Section~\ref{subec:PT} were obtained by using a $4\!\times\!4\!\times\!4$ supercell with $640$ atoms, lattice vectors of length $21.94$~\AA~ and angles between them of $55.83^\circ$. 
            For each temperature we simulated $10$ independent \texttt{NVT} runs (Langevin thermostat,  $\tau=100$~fs, $dt = 1$~fs). 
            We used the trained dipole model to evaluate the dipole for each step.

            The simulation with a time-dependent laser pulse, whose results are reported in Fig.~\ref{fig:LiNbO-phase-transition}c, were run with the same $4\!\times\!4\!\times\!4$ supercell as used above, and were run in the NVE ensemble, starting from a thermalized trajectory.  We used a time-step of 0.1fs to ensure stability of the integrator with a time-dependent Hamiltonian.
            For each pulse considered, we ran 16 independent trajectories starting from randomly sampled geometries equilibrated at 300~K.
            The average and standard deviation at varying time shown in Fig.~\ref{fig:LiNbO-phase-transition}c correspond to the statistical analysis over these independent trajectories.
            %

    \section*{Acknowledgments}
        E.S. thanks Federico Ernesto Mocchetti for the mathematical background regarding the topological aspects of the multi-valued \gls{ML} models and Shubham Sharma for discussions and help with training \texttt{MACE} models. We thank George Trenins for a critical read of the paper draft and insightful discussions.
        We thank Vasily Artemov for sending us a table with the experimental datapoints of the dielectric permittivity of water. We also thank Michael Fechner for insightful discussions about phonon dynamics in LiNbO$_3$. \rev{M.R. acknowledges funding by the European Research Council (ERC) Consolidator Grant (CoG) under the European Union’s Horizon Europe research and innovation programme (Grant agreement No. 101169761, Project Acronym QUADYMM).}

    \section*{Author Contributions}
        M.R. and E.S. conceptualized and designed the project. E.S., C.C. and M.R. discussed and refined the theory. E.S. implemented the models and conducted the calculations, with supervision from M.R.. E.S. and M.R. analysed the results. M.R., E.S. and C.C. wrote the manuscript.
        
       \section*{Competing Interests}
       The authors declare no competing interests.

    \bibliography{bibliography}
    
\end{document}


    \title{Electric-Field Driven Nuclear Dynamics of Liquids and Solids from a Multi-Valued Machine-Learned Dipolar Model\\Supplementary Information}

    \author{Elia Stocco}
    \affiliation{MPI for the Structure and Dynamics of Matter, Hamburg, Germany}
    
    \author{Christian Carbogno}
    \affiliation{Theory Department, Fritz Haber Institute of the MPS, Faradayweg 4-6, 14195 Berlin, Germany}

    \author{Mariana Rossi}
    \affiliation{MPI for the Structure and Dynamics of Matter, Hamburg, Germany}

    \maketitle

    \clearpage
    \section{Failure of a single-valued machine-learning model}
    
        \begin{figure}[H]
            \centering
            \begin{minipage}[t]{0.45\textwidth}
                \centering
                \includegraphics[width=\textwidth]{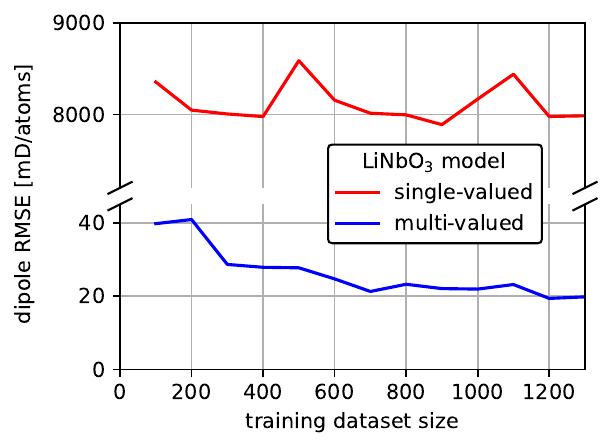}
            \end{minipage}
            \begin{minipage}[t]{0.45\textwidth}
                \centering
                \includegraphics[width=\textwidth]{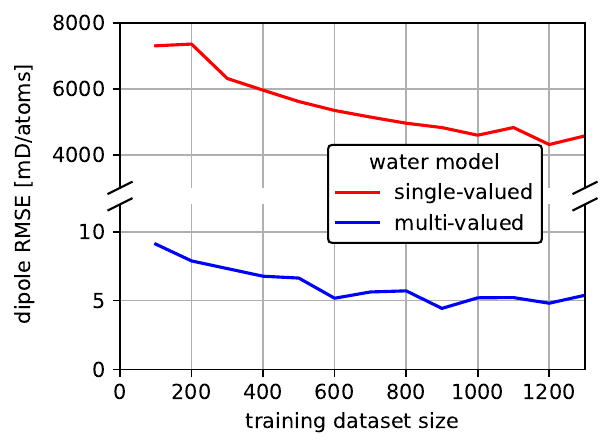}
            \end{minipage}
            \caption{
            Training of dipole ML models for polar systems including strongly out-of-equilibrium configurations.
            Specifically, the atomic structures used to evaluate the oxidation numbers (see example in Fig. S2 for water) have been added to both the train and test datasets, for LiNbO$_3$ (left) and water (right). The loss function (RMSE in milli-Debye) shows a clear failure of the single-valued model in both cases. We note that for water, removing the ``dissociated" configurations shown in Fig.~S2 would yield successful training for a single-valued model because the combined oxidation number $\mathcal{N}_I$ of the diffusing molecules is zero.}
            \label{fig:dipole-model-training}
        \end{figure}

        \begin{figure}[H]
            
                \centering
            \begin{minipage}[t]{\textwidth}
                \centering
                \includegraphics[width=\textwidth]{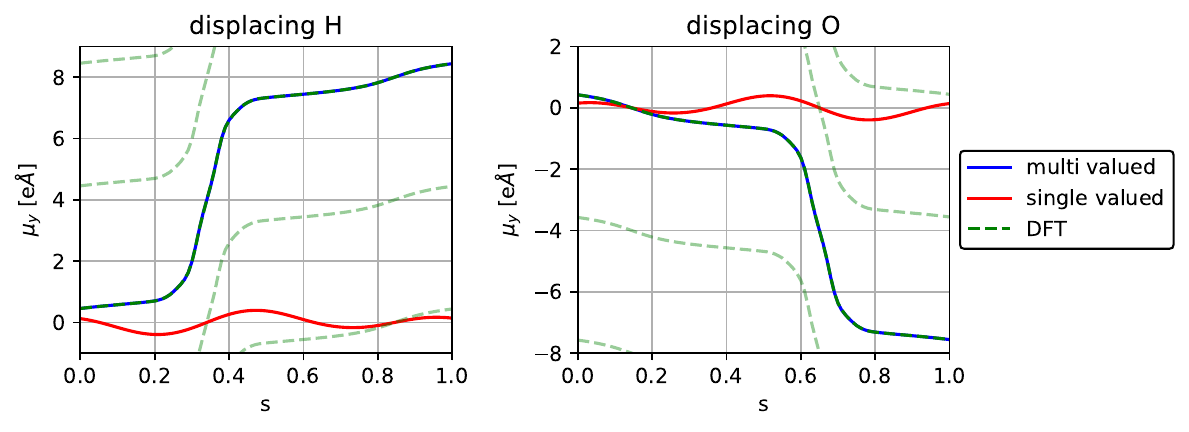}
            \end{minipage}
            
            \begin{minipage}[t]{\textwidth}
                \centering
                \includegraphics[width=0.8\textwidth]{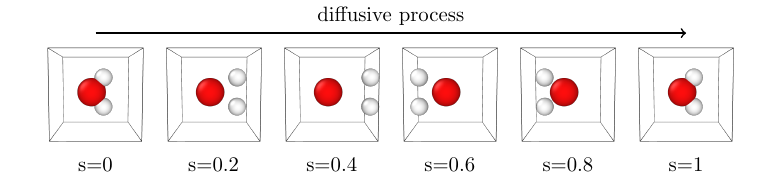}
            \end{minipage}
            \caption{Predicted dipole values (along the $y$-axis) of both the single- and multi-valued model of water along the path of Fig.1b and c of the main text.  The models used are the ones whose learning curves are shown in Fig.~S1. The left panel shows the trajectory where the oxygen atom is kept fixed while the hydrogen atoms diffuse to a neighboring periodic replica (also shown in the bottom plot), vice versa in the right panel.
            The DFT values are reported (in green) for different branches as well. 
            The multi-valued model is always in excellent agreement with the DFT data.
            The single-valued model roughly agrees with the data for $s\approx0$ (mod 1), where the coordinate $s$ is defined as in the main text, while for larger values of $s$ it completely diverges from the data by even missing the qualitative behavior and slope.}
            \label{fig:dipole-model-comparion}
        \end{figure}

    \clearpage
    \section{Frequency-dependent vibrational dielectric properties}

     \textbf{Supplementary Note 1}

            The following formula has been used to compute the infrared absorption spectrum $\alpha(\omega) n(\omega)$ of bulk water using the dipole time series~\cite{McQuarrie-Book}:
            \begin{align}
                \alpha\left(\omega\right) n \left(\omega\right) 
                = & \,
                \frac{\pi\omega}{3\hbar c \Omega \varepsilon_0}\left(1-e^{-\beta\hbar\omega}\right)
                C_{\bm{\mu}\bm{\mu}}\left(\omega\right) %
            \end{align}
            where $\alpha\left(\omega\right)$ is the Beer-Lambert absorption coefficient,  $n\left(\omega\right)$ is the  refractive index of the material, $\beta=1/k_{B}T$, $\hbar$ the reduced Planck constant, $c$ is the speed of light, $\varepsilon_0$ is the vacuum permittivity, $\Omega$ the system volume, and $C_{\bm{\mu}\bm{\mu}}\left(\omega\right)$ is the Fourier transform of the standard time correlation function $C_{\bm{\mu}\bm{\mu}}\left(t\right)$ of the dipole $\bm{\mu}$ with itself.
            By using the relation between the standard $C\left(\omega\right)$ and the Kubo transformed $\tilde{C}\left(\omega\right)$ time correlation function~\cite{KuboManolopoulos} one can re-express the previous expression as follows:
            \begin{align}
                \alpha\left(\omega\right) n \left(\omega\right) = & \, 
                \frac{\pi\beta\omega^2}{3 c \Omega \varepsilon_0}
                \tilde{C}_{\bm{\mu}\bm{\mu}}\left(\omega\right)
                \label{eq:IR-abs-spectrum-high-T}
            \end{align}
            where $\tilde{C}_{\bm{\mu}\bm{\mu}}\left(\omega\right)$ is approximated by using molecular dynamics with the classical time correlation function of the dipole with itself:            
            \begin{align}
                \tilde{C}_{\bm{\mu}\bm{\mu}}\left(\omega\right) \approx 
                \int\displaylimits_{-\infty}^{+\infty}dt\, e^{-i\omega t}\braket{\bm{\mu}\left(0\right)\cdot\bm{\mu}\left(t\right)}_\beta
            \end{align}

            The frequency-dependent dielectric susceptibility $\chi\left(\omega\right)$ is evaluated in a similar manner~\cite{McQuarrie-Book}:
            \begin{align}
                \chi\left(\omega\right) 
                = & \,
                - \frac{\beta}{3\varepsilon_0 \Omega} 
                \int\displaylimits_{0}^{+\infty}dt\, e^{-i\omega t}\braket{\bm{\mu}\left(0\right)\cdot\dot{\bm{\mu}}\left(t\right)}_\beta
            \end{align}
            
        \clearpage
        \section{Dipole model from periodic structures}
        
            \begin{figure}[H]
                \centering
                \includegraphics[width=\linewidth]{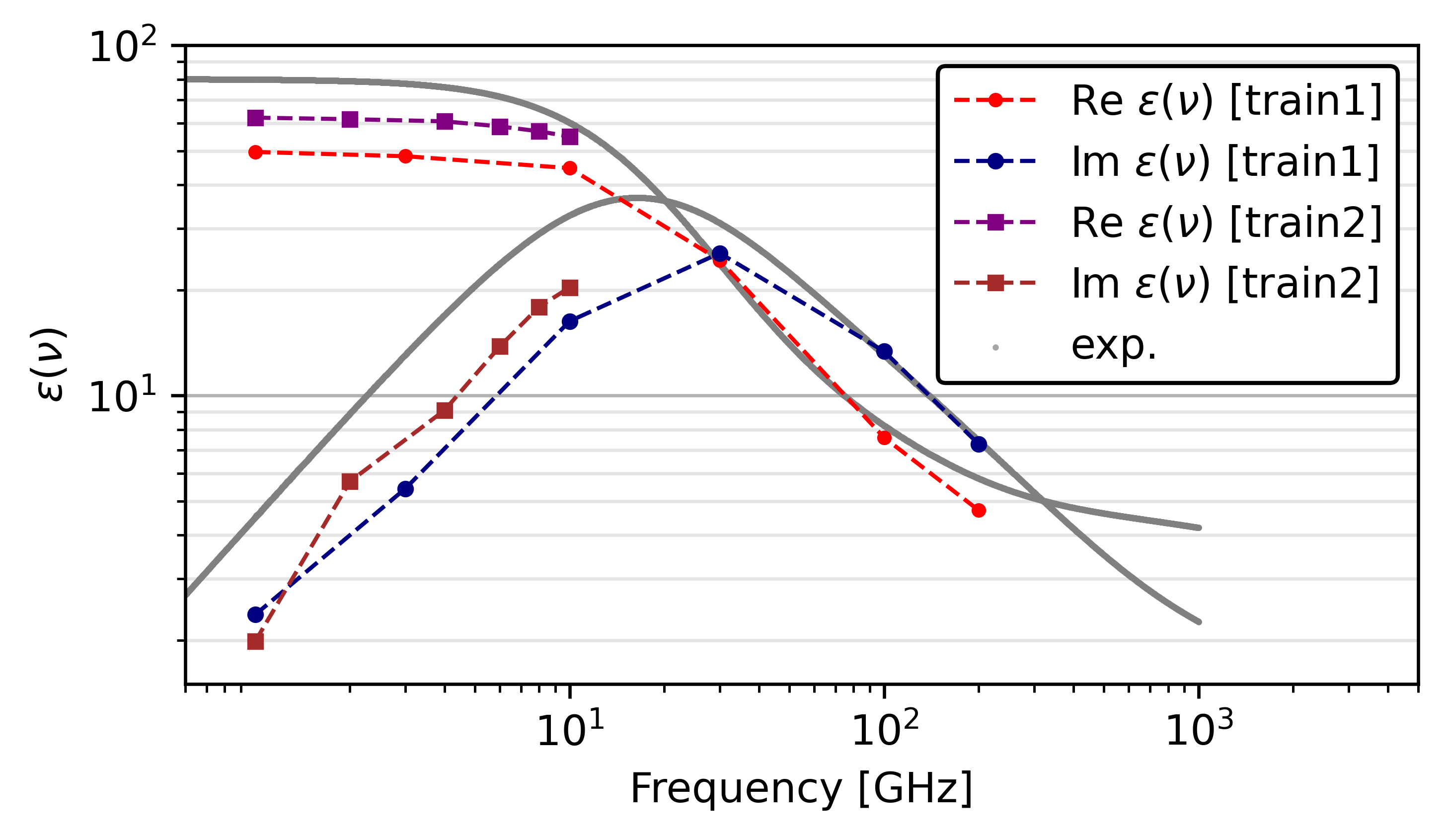}
                \caption{The figure presents the real and imaginary parts of the dielectric permittivity of liquid water at 300 K over the 1–135000 GHz frequency range, obtained using the non-equilibrium applied electric-field method (see the main text). The blue and red datasets (circles, train1) correspond to the results discussed in the main text, with the original training set for dipoles obtained from large water clusters. The purple and brown datasets (squares, train2) were obtained from a model trained on periodic water structures, instead of droplets. The dataset comprised of 1000 structures, each containing 32 water molecules at standard density. The dipoles were evaluated using Berry-phase polarization implementation in the FHI-aims code and the revPBE functional. The results indicate that data from periodic structures better captures the bulk regime and predicts a dielectric constant closer to the experimental value. Ramaining differences to experimental results are attributed to the functional used here (revPBE).}
            \end{figure}

        \clearpage
        \section{Quantum Infrared Spectrum of Liquid Water}

            \begin{figure}[H]
                \centering
                \includegraphics[width=\linewidth]{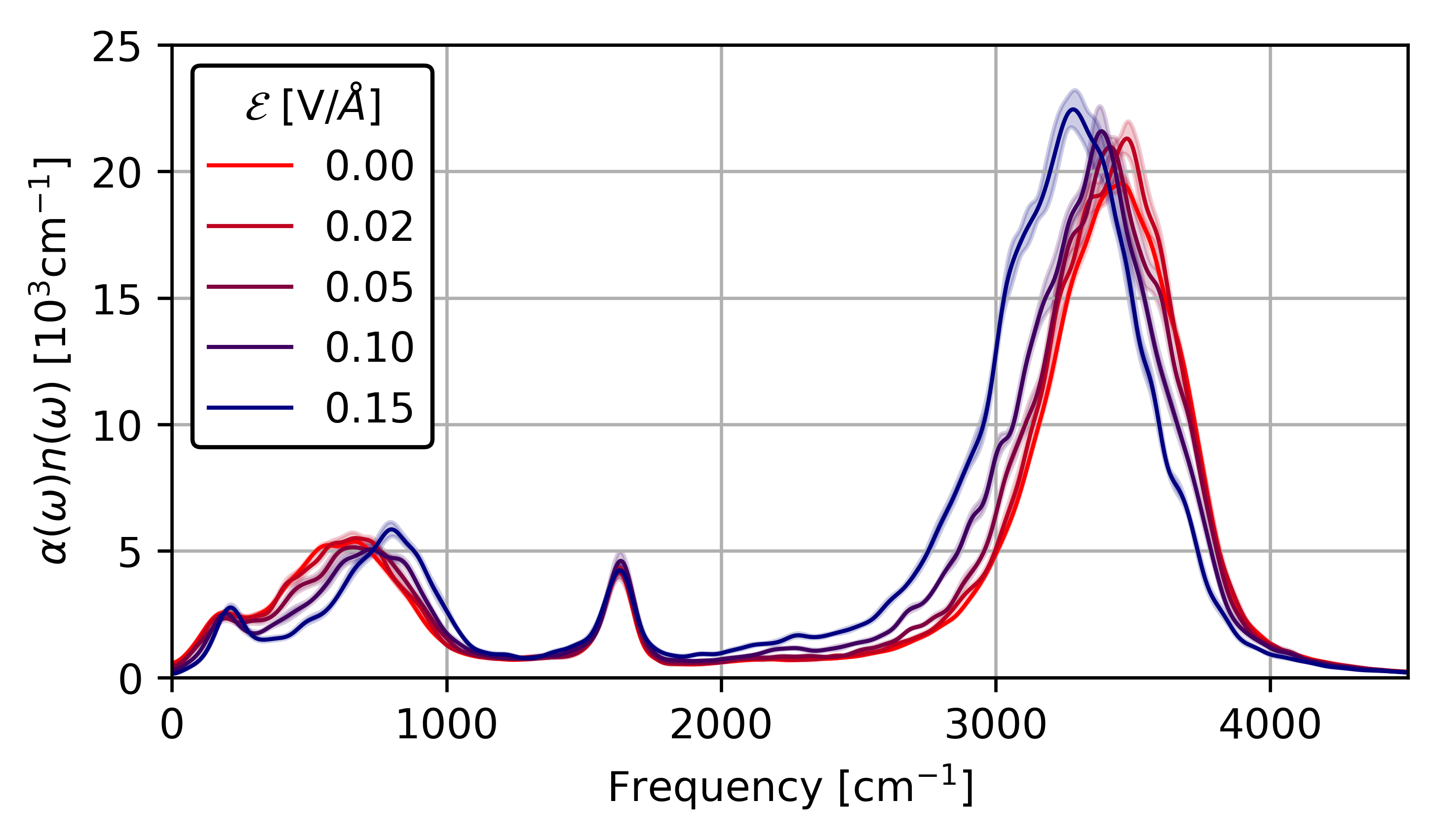}
                \caption{The infrared absorption spectrum of liquid water at varying electric field as obtained simulated with thermostatted ring-polymer molecular dynamics (TRPMD)~\cite{Rossi2014}, which includes nuclear quantum statistics. These spectra were obtained in a similar manner as done for the classical case shown in the main text. The values of the dipoles were obtained as the mean over the 32 beads. For each field intensity, 8 thermalized simulations of 10ps each were run with a time-step of 0.25~fs at 300~K using a global path integral Langevin equation thermostat (\texttt{pile\_g}) with $\tau=500$~fs and $\lambda=0.5$~\cite{Ceriotti2010}.}
            \end{figure}

        \clearpage
        \section{Phase transition of LiNbO$_3$}

            \begin{figure}[H]
                \centering
                \includegraphics[width=\textwidth]{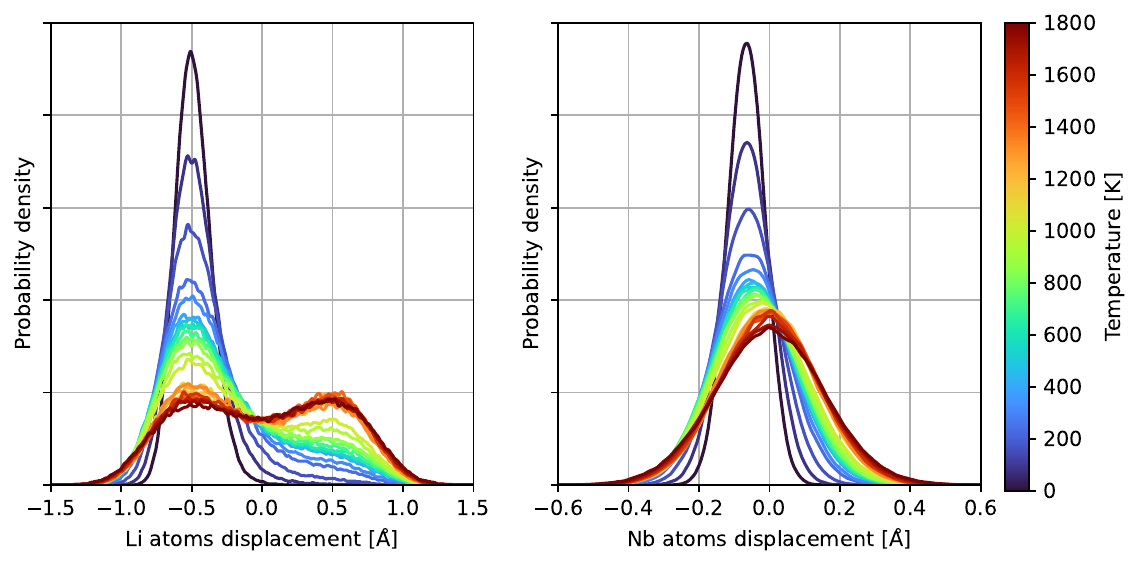}
                \caption{Normalized histograms of atomic displacements for Li and Nb atoms at various temperatures with respect to the paraelectric structure. The displacement distributions have been normalized by the maximum value and then smoothed using a Gaussian filter. These results are in agreement with what reported in Ref.~\cite{LNLT}, i.e.\ a bimodal distribution above the Curie temperature for Li, indicating a order-disorder phase transition, and a unimodal distribution for Nb, indicating a displacive phase transition.}
                \label{fig:LiNbO3-histogram}
            \end{figure}

        \clearpage
        \section{Fluence} \label{sec:fluence}
      
        \textbf{Supplementary Note 2}
    
            The fluence $F$~\cite{fluence} is defined as follows:
            \begin{align}
                F = & \, \frac{U}{A} = \frac{A}{A} \int_{-\infty}^{+\infty} dz \, u\left(z\right) 
            \end{align}
            where $U$ is the electromagnetic energy of the pulse, $A$ is the surface area of the sample, $u$ is the volumetric energy density (constant over the surface), and $z$ is the coordinate along the direction of propagation.
    
            If we change the integration domain from space to time, by using $c=z/t$ where $c$ is the speed of light in vacuum and $u=\varepsilon_0 \mathcal{E}^2\left(t\right)$, where $\varepsilon_0$ is the vacuum permittivity and $\mathcal{E}\left(t\right)$ is the electric field, we get
            \begin{align}
                F = \varepsilon_0 c \, \mathcal{I}
                 \qquad \text{with} \quad \mathcal{I} = \,  \int_{-\infty}^{+\infty} dt \, \mathcal{E}^2\left(t\right)
            \end{align}
            For the pulse adopted in this work, i.e.\ a plane wave of angular frequency $\omega$ with a gaussian envelope function of standard deviation $\sigma$, we can derive an explicit expression for $\mathcal{I}$:
            \begin{align}
                \mathcal{I} = & \,  
                    \int_{-\infty}^{+\infty} dt \, \mathcal{E}^2_{\rm max}
                    e^{- \frac{t^2}{\sigma^2}} \cos^2 \left(\omega t \right) \\
                    = & \, 
                    \int_{-\infty}^{+\infty} dt \, \mathcal{E}^2_{\rm max}
                    e^{- \frac{t^2}{\sigma^2}}
                    \left[ 
                    \frac{1+\cos \left(2\omega t \right)}{2}
                    \right] \\
                    = & \, 
                    \frac{\mathcal{E}^2_{\rm max}}{2}
                    \left[ \int_{-\infty}^{+\infty} dt \,
                    e^{- \frac{t^2}{\sigma^2}}        
                    + 
                    \int_{-\infty}^{+\infty} dt \, e^{- \frac{t^2}{\sigma^2}}  \cos \left(2\omega t\right)
                    \right] \\
                    = & \, 
                    \frac{\mathcal{E}^2_{\rm max}}{2}
                    \left[
                    \sigma \sqrt{\pi} +
                    \sigma \sqrt{\pi} 
                    e^{-\sigma^2\omega^2}
                    \right] \\
                    = & \, 
                    \mathcal{E}^2_{\rm max}
                    \frac{\sigma \sqrt{\pi}}{2}
                    \left(
                    1 +
                    e^{-\sigma^2\omega^2}
                    \right) 
            \end{align}
            where we have used to property of the gaussian integral and the Fourier transform of a gaussian.
            This lead to an explicit relation between the fluence $F$ and the parameters $\sigma,\omega$ and $\mathcal{E}_{\rm max}$ of the pulse:
            \begin{align}
                F = \varepsilon_0 c \, \mathcal{E}^2_{\rm max} \frac{\sigma \sqrt{\pi}}{2}
                    \left(
                    1 +
                    e^{-\sigma^2\omega^2}
                    \right) 
            \end{align}
    
            \begin{table}[H]
                \centering
                \begin{tabular}{c|c}
                    \toprule
                    $\mathcal{E}$ [V/\AA] & $F$ [mJ/cm$^2$] \\
                    \midrule
                    0.05 &  4.7  \\
                    0.15 &  42.3 \\
                    0.20 &  75.3 \\
                    0.25 & 117.6 \\
                    0.30 & 169.4 \\
                    \bottomrule
                \end{tabular}
                \caption{Conversion between the electric field intensity shown in the main text and the fluence for a laser pulse with $\sigma = 80$ fs (equal to a FWHM of $=188$~fs) and $\nu = 18$~THz.}
                \label{table:fluence}
            \end{table}

    \clearpage
    \section{Phonon driving of LiNbO$_3$}
    
        \begin{figure}[H]
            \centering
            \includegraphics[width=\textwidth]{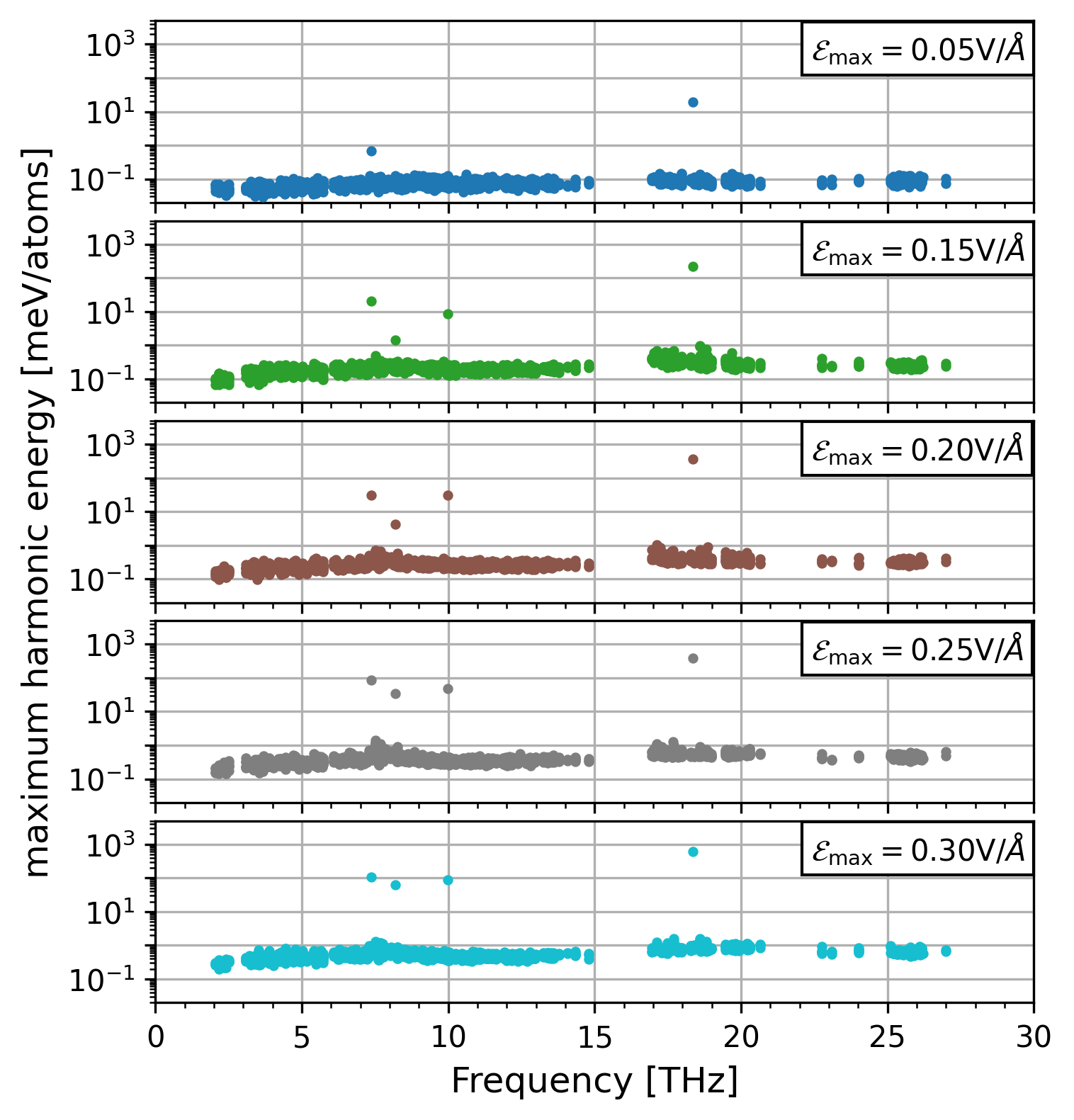}        
            \caption{Maximum harmonic energy assumed during the simulations by each the vibrational mode of the $4\!\times\!4\!\times\!4$ LiNbO$_3$ supercell. Only 4 modes, identified to be $\Gamma$-point modes and shown in Figure~\ref{fig:smodes} are excited (one directly by the pulse, and the other three by non-linear coupling). The projection procedure is described in~\ref{sec:phonons}.}
            \label{fig:mode-harm-energy}
        \end{figure}

        \begin{figure}[H]
            \centering
            \includegraphics[width=\textwidth]{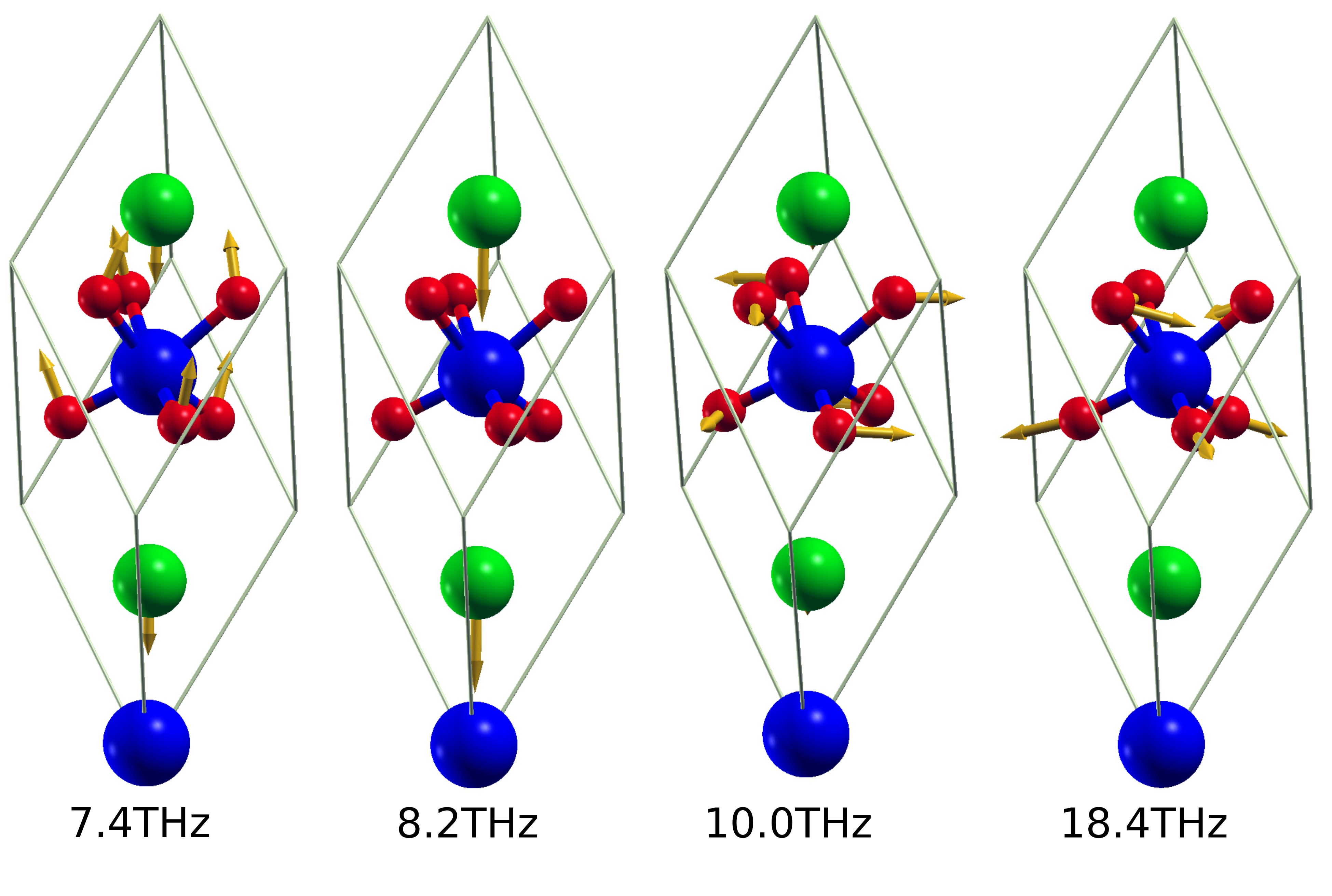}
            \caption{The 4 infrared active A$_1$ phonon modes of LiNbO$_3$ excited during the laser pulse, with their respective frequencies. The pulse was tuned at 18 THz. The frequencies of these modes were computed with \texttt{phonopy}~\cite{phonopy-phono3py-JPCM,phonopy-phono3py-JPSJ}. The modes at 7.4 THz and 18.4 THz correspond to $Q_{\rm P}$ and $Q_{\rm IR}$, respectively. The characters of the $Q_{\rm P}$ and $Q_{\rm IR}$ modes are in agreement with what reported in Ref.~\cite{MankowskiCavalleri2017}.\label{fig:smodes}}
        \end{figure}
        
        \begin{figure}[H]
            \centering
            \includegraphics[width=\textwidth]{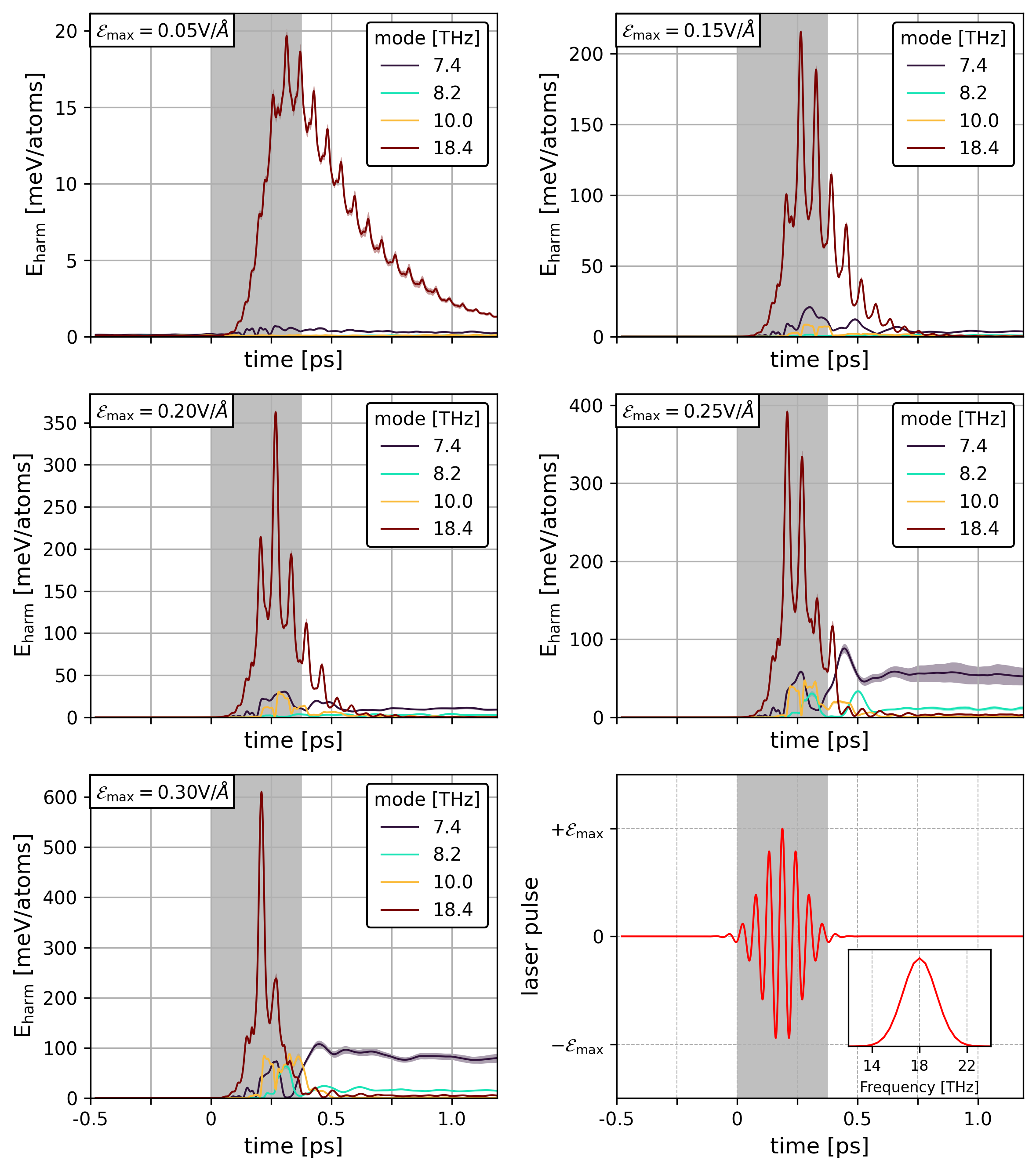}        
            \caption{Time evolution of the energy of the most excited modes due to the laser pulse. The pulse shape, described in the main text, is also shown in the last panel together with its power spectrum. The energies have been evaluated by projecting the trajectories along the harmonic normal modes, as exposed in~\ref{sec:phonons}.}
            \label{fig:mode-energy}
        \end{figure}

    \clearpage
    \section{Projection onto Phonon Modes}\label{sec:phonons}
    
       \textbf{Supplementary Note 3}
       
        This section describes the procedure for projecting molecular dynamics trajectories onto the phonon modes of a system to obtain a mode-resolved picture of the dynamics.
        Using a compact notation, nuclear displacements $\bm{q}$ and velocities $\bm{v}$ are written as:
        \begin{align}
            \bm{q} &= (q^1_x, q^1_y, q^1_z, \dots, q^{N_a}_x, q^{N_a}_y, q^{N_a}_z), \\
            \bm{v} &= (v^1_x, v^1_y, v^1_z, \dots, v^{N_a}_x, v^{N_a}_y, v^{N_a}_z).
        \end{align}
        In the harmonic approximation, the potential energy $U$ can be expressed as:
        \begin{align}
            \mathcal{H} = \frac{1}{2} \bm{v}^{\mathrm{t}} \cdot \doubleunderline{M} \cdot \bm{v} + \frac{1}{2} \bm{q}^{\mathrm{t}} \cdot \doubleunderline{\Phi} \cdot \bm{q},
        \end{align}
        where $\doubleunderline{M}$ is a diagonal mass matrix, and $\doubleunderline{\Phi}$ is the force constants matrix.
        The related eigenvalue problem is:
        \begin{align}
            \doubleunderline{D} \cdot \doubleunderline{\bm{\varepsilon}} = \doubleunderline{\bm{\varepsilon}} \cdot \doubleunderline{\Lambda},
        \end{align}
        where $\doubleunderline{D} = \doubleunderline{M}^{-1/2} \cdot \doubleunderline{\Phi} \cdot \doubleunderline{M}^{-1/2}$ is the dynamical matrix, and $\doubleunderline{\Lambda}$ contains the eigenvalues $\omega_n^2$, where $\omega_n$ are the angular frequencies of the normal modes.
        Displacements and velocities can be expressed in terms of normal modes:
        \begin{align}
            \bm{q}(t) &= \doubleunderline{M}^{-1/2} \cdot \bm{\varepsilon} \cdot \tilde{\bm{q}}(t), \\
            \bm{v}(t) &= \doubleunderline{M}^{-1/2} \cdot \bm{\varepsilon} \cdot \doubleunderline{\Lambda}^{1/2} \cdot \tilde{\bm{v}}(t),
        \end{align}
        where $\tilde{\bm{q}}(t)$ and $\tilde{\bm{v}}(t)$ are the mode coefficients. 
        These are computed from molecular dynamics trajectories by inverting the previous equations:
        \begin{align}
            \tilde{\bm{q}}(t) &= \doubleunderline{\bm{\varepsilon}}^{\mathrm{t}} \cdot \doubleunderline{M}^{1/2} \cdot \bm{q}(t), \\
            \tilde{\bm{v}}(t) &= \doubleunderline{\Lambda}^{-1/2} \cdot \doubleunderline{\bm{\varepsilon}}^{\mathrm{t}} \cdot \doubleunderline{M}^{1/2} \cdot \bm{v}(t).
        \end{align}
        Modes with $\omega \approx 0$, corresponding to rigid translations and rotations, can be discarded for numerical stability.
        The harmonic energy $E_n$ stored in each normal mode is given by:
        \begin{align}
            E_n = \frac{1}{2} \omega_n^2 \left( \tilde{v}^2_n + \tilde{q}^2_n \right)
        \end{align}
        This provides a mode-resolved decomposition of the dynamics. The same reasoning can be extended to off-$\Gamma$ modes, by considering the $\bm{k}$-dependence of all the previous quantities. A comprehensive exposition of the phonon theory can be found in Refs.~\cite{rigamonti2007structure,ashcroft1978solid}.

        \clearpage
        \section{Accuracies of the trained models}
            \begin{figure}[H]
                \centering
                \includegraphics[width=\textwidth]{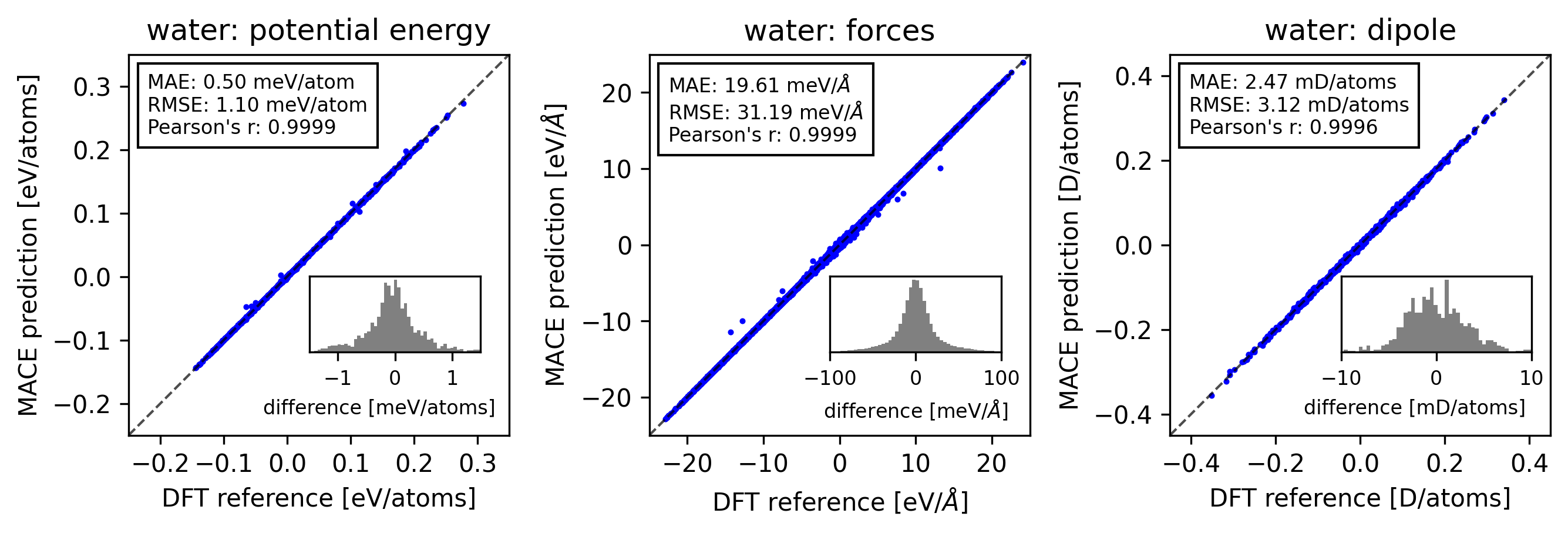}
                \caption{Correlation plots of MACE predictions with DFT reference values for liquid water on the training dataset for potential energy, forces, and on the test dataset for the dipole. For forces and dipoles, all components are plotted together. Inset histograms show the distribution of prediction errors for these quantities. We note that the outlier points in the forces correlation plot correspond to a single structure in the dataset, in which dissociation of water molecules is present.}
                \label{fig:water-correlation}
            \end{figure}
    
            \begin{figure}[H]
                \centering
                \includegraphics[width=\textwidth]{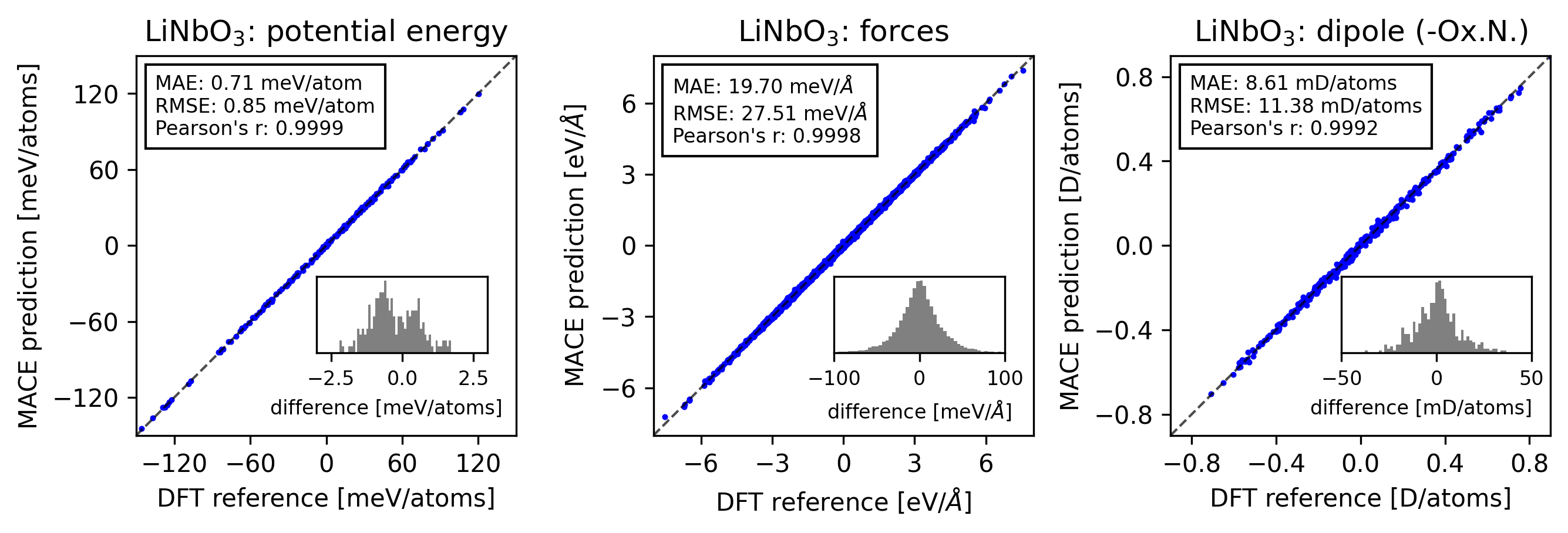}
                \caption{Correlation plots of MACE predictions with DFT reference values for LiNbO$_3$ on the test dataset for potential energy, forces, and dipole. For forces and dipoles all components are plotted together. Inset histograms show the distribution of prediction errors for these quantities. In the plots for the dipole, the oxidation number contribution has been subtracted for easier visualization.}
                \label{fig:LiNbO3-correlation}
            \end{figure}
    
            \begin{figure}[H]
                \centering
                \includegraphics[width=0.7\textwidth]{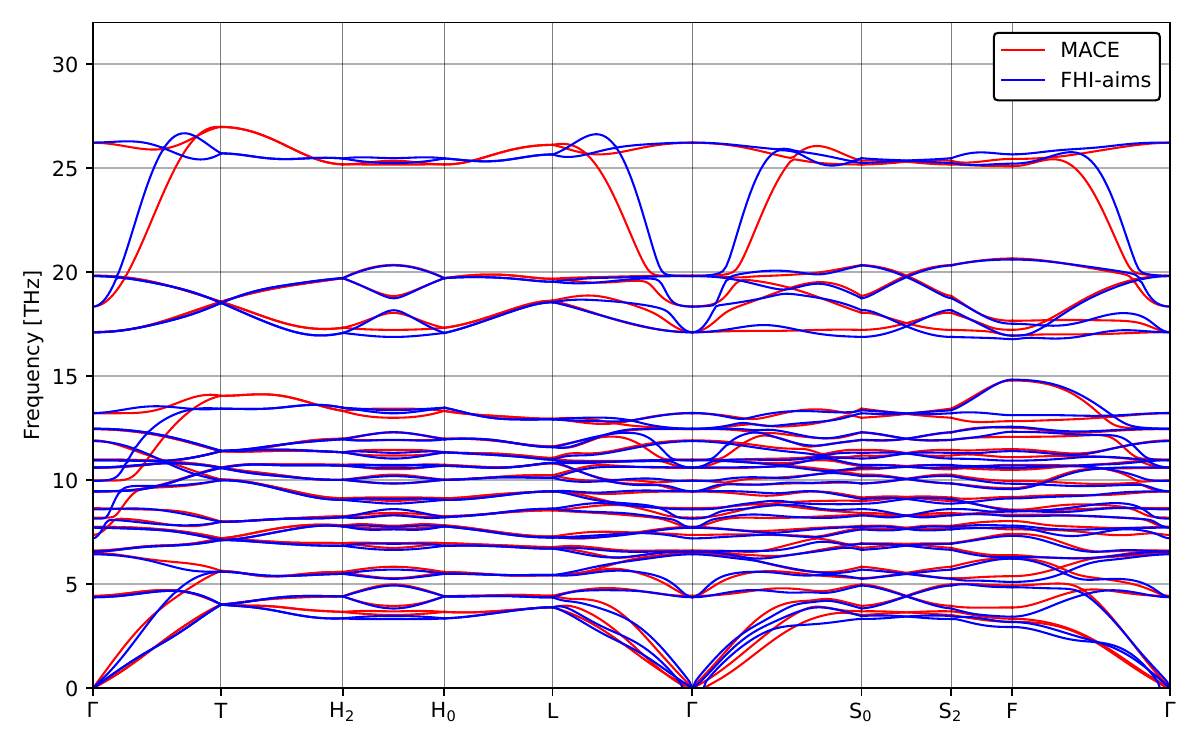}
                \caption{Comparison of the phonon bands of LiNbO$_3$ (without non-analytical corrections) between the trained \texttt{MACE} potential and \texttt{FHI-aims}. The phonons have been computed using a $4\!\times\!4\!\times\!4$ supercell and a $20\!\times\!20\!\times\!20$ mesh using \texttt{phonopy} \cite{phonopy-phono3py-JPCM,phonopy-phono3py-JPSJ}. The band paths along the Brillouin zone was generated using \texttt{seekpath}~\cite{HINUMA2017140}.}
            \end{figure}
        \clearpage
        \section{Dipole preprocessing}\label{appendix:dipole-preprocessing}
            
            \textbf{Supplementary Note 3}
            
                The values of the dipoles $\bm{\mu}$ as computed from \gls{DFT} need to be pre-processed in order to train any \gls{ML} model, since the values from different structures could lie on different branches.
                
                In order to fix the \textit{branch uncertainty}, we have followed the same procedure exposed in Ref.~\cite{Gigli2022}:
                \begin{mylist}
                    \item construct a toy model for the dipole $\bm{\mu}^{\rm ref}$ that will be used as a reference value at varying \glspl{nc} $\bm{R}$;
                    \item create a correlation plot of $\bm{\mu}^{\rm DFT}$ versus $\bm{\mu}^{\rm ref}$  for all the necessary structures;
                    \item Shift the clusters of points along the bundle of lines $y=nx$ with $n\in\mathbb{Z}$,
                    by the necessary number of quanta (along each $\bm{a}_{\alpha}$) such that all the point will lie along the same line.
                \end{mylist}
                If the chosen model is sufficiently descriptive, all $\bm{\mu}^{\rm DFT}$ will belong to the same branch after this procedure.

                There is a freedom of choice for the adopted toy model and for the chosen reference branch. We will explain it further in the following.
                Among the possible choices for a toy model we can cite:
                \begin{mylist}
                    \item partial charges model,
                    \item linear model,
                    \item a pre-trained \gls{ML} model.
                \end{mylist}
                A partial charges model is simply defined as $\bm{\mu}^{\rm PC}\left(\bm{R}\right) = \sum_I Q^I\bm{R}^I$ where $Q^I$ are real-valued atomic charges.
                Such a model, adopted in Refs. \cite{Jana2024}, is well suited for molecular and liquid systems, such as bulk water, and it's fairly easy to implement, depending only on few parameters. 
                A linear model, as the one implemented in Ref.~\cite{Gigli2022}, is defined as $\bm{\mu}^{\rm LM}\left(\bm{R}\right) = \bm{Z}^{*}_{\rm ref} \cdot \left( \bm{R} - \bm{R}_{\rm ref}\right)$, and it's suitable for solid systems where it's easy to identify a reference structure $\bm{R}_{\rm ref}$ (usually a paraelectric ground state).
                Such a model is effective for solids because all the considered structures (at least for low enough temperature) can be assumed close enough to $\bm{R}_{\rm ref}$, not showing any pronounced non-linearities.
                However, this approach depends on a number of parameters which are linear in the system size and cannot be transferred ``out-of-the-box'' to cells with a different number of atoms, since $Z^*_{\rm ref}$ and $\bm{R}_{\rm ref}$ are ``cell-dependent''.
                A pre-trained \gls{ML} model for the dipole has the advantage to grasp nonlinearities in the dipole, but of course it can be used only at a second stage while at a first stage one of the previous model has to be adopted. 
                However, such a model allows to generate a self-consistent workflow able to correct data that were placed in the wrong branch on a previous stage.  

                The correction of the data by using one of the previously mentioned method will move all the data to the same branch, which is determined by the chosen reference model.
                However, as already recognized in Ref. \cite{Gigli2022}, even though from the physical point of view any branch is equivalent, from a practical point of view this is not the case when working with $E\left(3\right)$-equivariant models.
                In a $E\left(3\right)$-equivariant framework, the dipole $\bm{\mu}$ is odd with respect to\ \gls{nc} $\bm{R}$, i.e.\ $\bm{\mu}\left(-\bm{R}\right) = - \bm{\mu}\left(\bm{R}\right)$ (where a rotation of the lattice vector $\bm{a}_{\alpha}$ is understood as well).
                This means that we can not shift the values of the dipoles by any $e\bm{a} \cdot \bm{n}$ since:
                \begin{align}
                    \bm{\mu}\left(\bm{R}\right)  \longrightarrow \bm{\mu}'\left(\bm{R}\right) = & \, \bm{\mu}\left(\bm{R}\right) + e\bm{a} \cdot \bm{n}  \notag \\
                    \bm{\mu}\left(-\bm{R}\right)  \longrightarrow \bm{\mu}'\left(-\bm{R}\right) = & \,  \bm{\mu}\left(-\bm{R}\right) + e\bm{a} \cdot \bm{n} = -\bm{\mu}\left(\bm{R}\right) + e\bm{a} \cdot \bm{n} \notag \\
                    \bm{\mu}'\left(\bm{R}\right)  \neq & \,  - \bm{\mu}'\left(-\bm{R}\right)
                \end{align}
                This means that the use of a $E\left(3\right)$-equivariant \gls{ML} models \textit{imposes} to chose the only branch where equivariance can be satisfied.
                A possible way to achieve this goal is to use a reference model which is explicitly $E\left(3\right)$-equivariant, such as a point-charges model (being linear in $\bm{R}$).
                In this paper we have used at a first stage a point charges model where the partial-charges $Q$ were set equal to the oxidation numbers $\mathcal{N}_I$ for LiNbO$_3$ and half of their value for bulk water, and on a second stage the trained \gls{ML} model in a self-consistent manner in order to ensure that all the data were placed in the correct branch.

        \clearpage
\renewcommand{\refname}{Supplementary References}
        \bibliography{bibliography}